\documentclass[12pt]{article}
\pdfoutput=1

\usepackage[utf8]{inputenc}

\usepackage{amsmath,amssymb,amsfonts,amscd,mathrsfs}
\usepackage{xcolor}
\definecolor{darkblue}{rgb}{0.1,0.1,.7}
\usepackage[colorlinks, linkcolor=darkblue, citecolor=darkblue, urlcolor=darkblue, linktocpage]{hyperref} 
\usepackage[square, comma, compress,numbers]{natbib}
\usepackage[]{graphicx}

\usepackage{geometry}
\usepackage[margin=10pt,font=small,labelfont=bf]{caption}
\usepackage{ifthen}
\usepackage{tikz}
\usepackage{subcaption}
\usepackage{booktabs,multirow}
\usepackage{hhline}
\usepackage{slashed}
\usepackage{hyperref}
\usepackage{bm}

\usepackage{ragged2e}
\usepackage{array}
\newcolumntype{L}[1]{>{\raggedright\let\newline\\\arraybackslash\hspace{0pt}}m{#1}}
\newcolumntype{C}[1]{>{\centering\let\newline\\\arraybackslash\hspace{0pt}}m{#1}}
\newcolumntype{R}[1]{>{\raggedleft\let\newline\\\arraybackslash\hspace{0pt}}m{#1}}

\usepackage{booktabs}

\def\bit{\begin{itemize}}
\def\eit{\end{itemize}}
\def\baa{\begin{array}}
\def\eaa{\end{array}}

\def\eps{\epsilon}

\newcommand{\beq}{\begin{equation}} 
\newcommand{\eeq}{\end{equation}}

\newcommand{\diffop}[2]{\ifthenelse{\equal{#2}{1}}{\frac{\mrm{d}}{\mrm{d} #1}}{\frac{\mrm{d}^#2}{\mrm{d} #1^#2}}}

\newcommand{\mrm}[1]{{\mathrm #1}}

\usepackage[normalem]{ulem}

\newcommand{\be}{\begin{equation}}
\newcommand{\ee}{\end{equation}}
\newcommand{\bea}{\begin{eqnarray}}
\newcommand{\eea}{\end{eqnarray}}

\usepackage{hyperref}
\newlength{\dhatheight}

\numberwithin{equation}{section}
\begin{document}

\vspace*{-.6in} \thispagestyle{empty}
\begin{flushright}
 SISSA 01/2019/FISI
\end{flushright}
\vspace{1cm} {\Large
\begin{center}
{\bf 
{Precision diboson measurements at hadron colliders}
}
\end{center}}
\vspace{1cm}

\begin{center}
{ \bf A. Azatov , D. Barducci  ,  E. Venturini} 

{
 SISSA and INFN sezione di Trieste, Via Bonomea 265; I-34137 Trieste, Italy
 
}
\vspace{1cm}
\end{center}

\vspace{4mm}

\begin{abstract}
We discuss the measurements of the anomalous triple gauge couplings at Large Hadron Collider focusing on the contribution of the ${\cal O}_{3W}$ and ${\cal O}_{3\tilde W}$ operators. 
These deviations were known to be particularly hard to measure due to their suppressed interference with the SM amplitudes in the inclusive processes, leading to  approximate flat directions in the space of these Wilson coefficients. We present the prospects for the measurements of these interactions at HL-LHC and HE-LHC  using exclusive variables sensitive to the interference terms and taking carefully into account effects appearing due to NLO QCD corrections.
 \end{abstract}
\vspace{.2in}
\vspace{.3in}
\hspace{0.7cm} 

\vfill
\noindent\line(1,0){188}
{\scriptsize{ \\ E-mail:
\texttt{\href{mailto:aleksandr.azatov@sissa.it}{aleksandr.azatov@sissa.it}},
\texttt{\href{mailto:daniele.barducci@sissa.it}{daniele.barducci@sissa.it}}, 
\texttt{\href{mailto:eventuri@sissa.it}{eventuri@sissa.it}}
}}

\newpage

\setcounter{tocdepth}{2}

{
\tableofcontents
}

\section{Introduction}
The Standard Model (SM) is an exceptionally powerful theory in describing most of the observed phenomena in particle physics. The experimental measurements at the Large Hadron Collider (LHC) following the discovery of the Higgs boson~\cite{Aad:2012tfa,Chatrchyan:2012xdj} continue to confirm it to be the valid theory for the larger and larger scales of energy. 
At the same time we know that the SM has to be just an effective field theory, since it cannot 
provide explanations to various experimental observations such as the existence of neutrino 
masses and oscillations, the existence of dark matter and the large baryon asymmetry present in the Universe. Moreover, fine tuning arguments such as the naturalness of the electroweak (EW) scale, the large hierarchy among the Yukawa couplings of the SM fermions and the non observation of charge-parity (CP) violation effects in strong interactions also seem  to suggest that an ultraviolet completion of this theory is needed.
In particular, in the SM the Higgs mass parameter responsible for EW symmetry breaking is quadratically sensitive to any heavy new physics (NP) contribution. This hints to a relative low energy scale where new dynamic degrees of freedom should be present, unless one is willing to accept a large level of fine tuning in the EW sector. This paradigm has motivated in the last decades a huge experimental effort for the direct search for NP at and beyond the EW scale.
The null results so far obtained have however almost ruled out the vanilla new physics models, except in some corners of their parameter space.

In view of this, precision studies of all the possibile deformations from the SM due to new states not directly accessible at  current collider energies become a crucial task for the present and future experimental program. The language of the SM Effective Field Theory (SMEFT) provides a well defined organizing principle for characterizing the various deviations from the SM Lagrangian,
given the (at least moderate) mass gap existing between the EW scale and the NP scale $\Lambda$.
 As well known in this language the new interactions are expressed as a series of higher dimensional operators so that the effective Lagrangian can be written as
\be\label{eq:eft}
{\cal L}_{{\rm SMEFT}}={\cal L}_{{\rm{SM}}}+ {\cal L }_6+...
\ee
where
${\cal L }_i = \sum_i \frac{c_i O_i}{\Lambda^{i-4}}$
and $c_i$ are the Wilson Coefficients of the operators $O_i$, built out from SM fields. The expression in Eq.~\eqref{eq:eft} is valid under the assumption of lepton  number conservation, so that the first corrections to the SM only appear at the order of dimension six operators, order for which a complete basis has been identified long ago in~\cite{Buchmuller:1985jz,Grzadkowski:2010es}. One of the main goals of the current and High-Luminosity program of the LHC (HL-LHC), as well as of future High-Energy options (HE-LHC), is the precise determination of the $c_i$ coefficients. The main objective of this paper is the study for the measurement of two dimension six operators affecting the triple gauge coupling among EW gauge bosons, namely
\be
\label{eq:operators}
O_{3W}=- \frac{1}{\Lambda^2}\frac{g}{3!}{ \eps_{abc} W^{a,\mu\nu}W_{\nu\lambda}^bW^{c\lambda}_{~~\mu}}
\quad\quad\quad
O_{3\tilde W}=- \frac{1}{\Lambda^2}\frac{g}{3!}{ \eps_{abc} \tilde W^{a,\mu\nu}W_{\nu\lambda}^bW^{c\lambda}_{~~\mu}}
\ee
where $g$ and $W_{\mu\nu}$ are the $SU(2)_L$ gauge coupling constant and the field strength tensor and  $\tilde W^{\mu\nu}$ its dual,   $\tilde W^{\mu\nu}=\frac{1}{2} \epsilon^{\alpha\beta\mu\nu}W_{\alpha\beta}$. 
It is well known that the measurement of the Wilson coefficient of the two operators of Eq.~\eqref{eq:operators} is extremely challenging since the interference between the SM and NP contributions to 
diboson production in $2\to 2$ scattering is suppressed in the high-energy regime as a consequence of certain helicity selection rules~\cite{Dixon:1993xd,Azatov:2016sqh}. This makes it hard to precisely determine the magnitude of the $c_{3W}$ and $\tilde c_{3 W}$ Wilson coefficients, as well as to measure their sign and to differentiate amongst their two different contributions to the scattering amplitudes. Based on the fact that the helicity selection rules of~\cite{Azatov:2016sqh} are only valid for $2\to 2$ scattering, various observables built out from the decay products of the diboson final states have been recently proposed~\cite{Azatov:2017kzw,Panico:2017frx}. These observables help to overcome the non interference problem ensuring a larger sensitivity to the Wilson coefficient of the operators of Eq.~\eqref{eq:operators}.

In this paper we update previous analyses by considering both the $pp\to WZ$ and $p p \to W\gamma$ diboson processes with the inclusion of the $O_{3\tilde W}$ operator and carefully treating QCD next-to-leading-order (NLO) effects, which we find to be important since they partially restore the interference between the SM and BSM amplitudes (for the previous studies of the NLO effects in the presence of the higher dimensional operators see \cite{Campanario:2016jbu,Baglio:2017bfe,Chiesa:2018lcs}).

By closely following existing experimental analyses targeting diboson final states, we provide combined bounds in the $c_{3W}-\tilde{c}_{3W}$ plane showing the potentiality of the HL- and HE-LHC options in testing these higher dimensional operators. 
Interestingly we find that some of the selection cuts which are necessary to suppress reducible QCD background processes automatically lead to a partial restoration of the interference also at LO, an effect extremely relevant for experimental analyses and which was overlooked in the previous literature. 

 We also compare our findings for the $O_{3\tilde W}$ operator with the limits arising from the non observation of neutron and electron electric dipole moments (EDM). We find that the HL-LHC sensitivity on the CP odd operator becomes stronger than the bounds from neutron EDM but not than the ones from the electron EDM, which are one order of magnitude stronger.

The paper is organized as follows. In Sec.~\ref{sec:inter} we give a review of the methods recently proposed in the literature to restore the interference of the operators of Eq.~\eqref{eq:operators}. We then present our analyses for the $WZ$ and $W\gamma$ processes in Sec.~\ref{sec:WZ} and Sec.~\ref{sec:WA}. Limits from EDMs are discussed in Sec.~\ref{sec:EDM} while prospect for the HE option of the LHC at 27 TeV are shown in Sec.~\ref{sec:LHC27}.  We conclude in Sec.~\ref{sec:concl}.


\section{Interference suppression and its restoration}\label{sec:inter}

In this section we will review the main results regarding the interference suppression from the helicity selection rules~\cite{Azatov:2016sqh} and the possible strategies to overcome it recently proposed in~\cite{ Azatov:2017kzw,Panico:2017frx}. The reader familiar with the topic can directly skip to the next Section. 

Generically, the scattering cross section for any $2\to 2$ process in the presence of higher dimensional beyond the SM (BSM) operators can be written as
\bea
\label{eq:sigtt}
\sigma\sim \frac{g_{\text{SM}}^4}{E^2}\bigg[ \overbrace{ \left(a^{{\rm{SM}}}_0 +a^{{\rm{SM}}}_1\frac{M^2}{E^2}+...\right)}^\text{SM$^2$}
  & +\overbrace{\frac{E^2}{\Lambda^2}  \left(a^{{\rm{int}}}_0 +a^{{\rm{int}}}_1\frac{M^2}{E^2}+...\right)}^\text{BSM$_6\times\,$SM}
       +\overbrace{ \frac{E^4}{\Lambda^4}\left( a^{{\rm{BSM}}}_0 +a^{{\rm{BSM}}}_1\frac{M^2}{E^2}+...\right)}^\text{BSM$_6$$^2$}  \bigg]\, ,\nonumber\\ 
\eea
where $E$ is the typical energy of the scattering process, $M$ is the mass of the SM particles and ellipses stand for the smaller terms in the $\left(\frac{M^2}{E^2}\right)$ expansion
The interference terms between the SM and BSM as well as a pure BSM terms are indicated explicitly.
In the high energy limit $E\gg M$ the leading contribution comes from the $a_{0}^{{\rm{SM,int,BSM}}}$ terms in the brackets
  corresponding to the zero mass limit of the SM particles.
In~\cite{Azatov:2016sqh} it was shown that $a_0^{{\rm{int}}}$ (the leading contribution to the interference term)
is equal to zero for all of the processes 
containing  transversely polarized vector bosons.
This effect comes from the fact that the SM and NP amplitudes contain  transverse vector bosons in the different helicity eigenstates, for which the interference vanishes. 
Dramatically, this interference suppressions implies that the high energy  
measurements of the Wilson coefficients will not benefit from the usual growth of 
the amplitudes with the energy expected from dimension six operators.  
This negatively affects the possibilities of high-energy hadron colliders, where the  strongest bounds can usually be obtained by exploiting the relative enhancement of the NP contribution compared to the SM one in the high energy distribution tails \cite{Baur:1994aj,Falkowski:2016cxu,Berthier:2016tkq,Butter:2016cvz,
Dumont:2013wma,Ellis:2014dva,Franceschini:2017xkh,Grojean:2018dqj,
Biekotter:2018rhp,Liu:2018pkg,Alves:2018nof,Almeida:2018cld}.

\subsection{Modulation from azimuthal angles: ideal case}
For concreteness let us consider the process $qq\to V_T V_T$, where $V=W^\pm,Z,\gamma$ and  we will always work in the high energy limit, $E\gg m_V$.
In the SM then the only amplitudes that will be generated at leading order in energy are $A_{{\rm SM}}(q \bar q \to V_{T,\pm} V_{T,\mp})$, where the helicities of the final state 
vector bosons are explicitly indicated. At the same time the dimension six operators in 
Eq.~\eqref{eq:operators} generate only the amplitudes  $A_{{\rm BSM}}(q \bar q 
\to V_{T,\pm} V_{T,\pm})$. Clearly, there is no interference between the BSM and SM contributions. This is the core of the above mentioned helicity selection rules. %
However note that at least one of  the vector bosons in the final sate is  not stable. Hence  the physical process is not a $2\to2 $ but instead a $2\to 3$ or $2\to 4$ scattering. For simplicity let us consider the case of $q \bar q \to W_T \gamma$ with a leptonically decaying $W$. The differential cross section can then be schematically written as
\be
\frac{{\rm{d}}\sigma(q\bar q \rightarrow \gamma_{+} l_- \bar \nu_+)}{{\rm{d}}\text{LIPS} } =
\frac{1}{2 s}  \frac{ \left|\sum_i  
	(  {\cal M}^\text{SM}_{q\bar q \rightarrow \gamma_{+}W_{i}}+
	{\cal M}^\text{BSM}_{q\bar q \rightarrow \gamma_{+}W_{i}}
	)
	{\cal M}_{W_{i}\to  l_- \bar \nu_+}   \right|^2}{(k_W^2-m_W^2)^2+m_Z^2\Gamma_W^2} 
\,  , \label{xsecan}
\ee
where the  sum runs over the intermediate $W$ polarizations, with the $W$
assumed to be on-shell, and where ${\rm{d}}\text{LIPS}\equiv (2\pi)^4\delta^4(\sum p_i -p_f) \prod_i \frac{d^3 p_i}{2 E_i(2\pi)^3}$
 is the Lorentz Invariant Phase Space. %
In the narrow width approximation the leading contribution to the interference, {\emph{i.e.}} the cross term $\text{SM}\times\text{BSM}$ in Eq.~\eqref{xsecan},  is given by:
\be
\frac{\pi}{2 s}  \frac{\delta (s-m_W^2)}{\Gamma_W m_W}   
{\cal M}^\text{SM}_{q\bar q \rightarrow \gamma_{+}W_{T_-}} \left(  {\cal M}^\text{BSM}_{q\bar q \rightarrow \gamma_{+}W_{T_+}}  \right)^* {\cal M}_{W_{T_-}\to  l_- \bar \nu_+} 
{\cal M}_{W_{T_+}\to   l_- \bar \nu_+}^*
+ h.c., \label{dxsc1}
\ee
where we have ignored the contributions to the longitudinal polarizations in the SM. 
A simple  calculation shows that 
\bea
\label{eq:prodamp}
 {\cal M}_{W_{T_-}\to  l_- \bar \nu_+} 
{\cal M}_{W_{T_+}\to   l_- \bar \nu_+}^*\propto e^{- 2 i\phi}
\eea
where $\phi$ is the angle spanned by the plane of the $W$ decay products and the $W\gamma$ scattering plane.  
As explicitly shown in~\cite{Panico:2017frx}, the phase of the expression
${\cal M}^\text{SM}_{q\bar q \rightarrow \gamma_{+}W_{T_-}} \left(  {\cal M}^\text{BSM}_{q\bar q \rightarrow \gamma_{+}W_{T_+}}  \right)^*$ can be identified 
using the optical theorem and its properties under CP transformations. Let's consider an 
arbitrary amplitude  $A(a\to b)$. Then the optical theorem (if there are no strong phases, {\emph{i.e.}} contributions of   nearly on-shell particles) fixes
\bea\label{eq:ab1}
A(a\to b)= A^*(b\to a).
\eea
At the same time the transformation under CP implies
\bea\label{eq:ab2}
A(a\to b)= \eta_{CP} A(b\to a)
\eea
where  $\eta_{CP}=1(-1)$ for interactions respecting (violating)  CP symmetry. By combining Eq.~\eqref{eq:ab1} and Eq.~\eqref{eq:ab2} we can infer
\bea
A(a\to b)^*= \eta_{CP} A(a\to b).
\eea
Applying this result to the $qq\to W\gamma$ process we obtain that
\bea
\label{eq:cprelation}
{\cal M}^\text{SM}_{q\bar q \rightarrow \gamma_{+}W_{T_-}} \left(  {\cal M}^\text{BSM}_{q\bar q \rightarrow \gamma_{+}W_{T_+}}  \right)^*= \eta_{CP}(\rm{BSM})
\left[
{\cal M}^\text{SM}_{q\bar q \rightarrow \gamma_{+}W_{T_-}} \left(  {\cal M}^\text{BSM}_{q\bar q \rightarrow \gamma_{+}W_{T_+}}  \right)^*\right]^*.
\eea
By using the results in the Eq.~\eqref{eq:cprelation} and Eq.~\eqref{eq:prodamp}
 we can see that the differential cross sections  from the $\text{SM}\times\text{BSM}$ interference arising from the insertion of the $O_{3W}$ and $O_{3\tilde W}$ operators have the following form
\bea
\label{eq:idealmodWA}
O_{3 W}: ~~~ {\cal M}_{W_{T_-}\to  l_- \bar \nu_+} 
{\cal M}_{W_{T_+}\to   l_- \bar \nu_+}^* + h.c.\propto \cos(2\phi_W)\nonumber\\
O_{3 \tilde W}: ~~~ {\cal M}_{W_{T_-}\to  l_- \bar \nu_+} 
{\cal M}_{W_{T_+}\to   l_- \bar \nu_+}^* - h.c. \propto \sin(2\phi_W)\nonumber.\\
\eea
Similar arguments can be applied to the case of $WZ$ production. There, since only one pair of the intermediate vector bosons have opposite helicities,  the modulation factorizes into a sum of two independent terms and reads
\bea
\label{eq:idealmodWZ}
&&O_{3 W}: ~~~\propto \cos(2\phi_W)+\cos (2\phi_Z)\nonumber\\
&&O_{3 \tilde W}: ~~~\propto \sin(2\phi_W)+\sin (2\phi_Z).
\eea

The take home message is that by exploiting the modulations of Eq.~\eqref{eq:idealmodWA} and Eq.~\eqref{eq:idealmodWZ} is possible to increase the precision on the determination of the Wilson coefficients associated with the $O_{3W}$ and $O_{3\tilde W}$ operators by overcoming the suppression of the interference terms of the cross section, suppression that is recovered with no ambiguity by performing a complete integration over the $\phi_i$ angles.

\subsection{Modulation from azimuthal angles: real case}

In the previous Section we have discussed an ideal situation, assuming that the azimuthal angles between the plane spanned by the vector bosons decaying products and the scattering plane can be exactly determined. However the azimuthal angle determination suffers from a twofold degeneracy as pointed out in~\cite{Panico:2017frx}. 
Let us recall the definitions of the $\phi$ angles which can be used experimentally. First we define two normals
\bea
\label{eq:planes}
&&\hat{n}_{{\rm{decay}} }^{i}\parallel \vec{p}_{l^{i},+} \times\vec{p}_{l^{i},-} \nonumber\\
&&\hat{n}_{{\rm{scat.}} }^{i} \parallel \hat{z}_{{\rm{lab.}}} \times\vec{p}_{V^{i}} 
\eea
where the index $i$ refers to the first or the second vector boson,  $l^{i}$
are the leptons from its decay and $\pm$ indicate the lepton  helicities. The azimuthal angle $\phi$ between the two planes orthogonal to the normals is thus defined as
\bea
\label{eq:angledef}
\phi_V=\hbox{sign}\left[ (\hat{n}_{{\rm{scat.}}}^{i}\times\hat{n}_{{\rm{decay}}}^{i})\cdotp\vec{p}_{V^{i}}\right] \arccos (\hat{n}_{{\rm{scat.}}}^{i}\cdotp\hat{n}_{{\rm{decay}}}^{i}).
\eea

\begin{figure}[t!]
\centering
\includegraphics[scale=0.7]{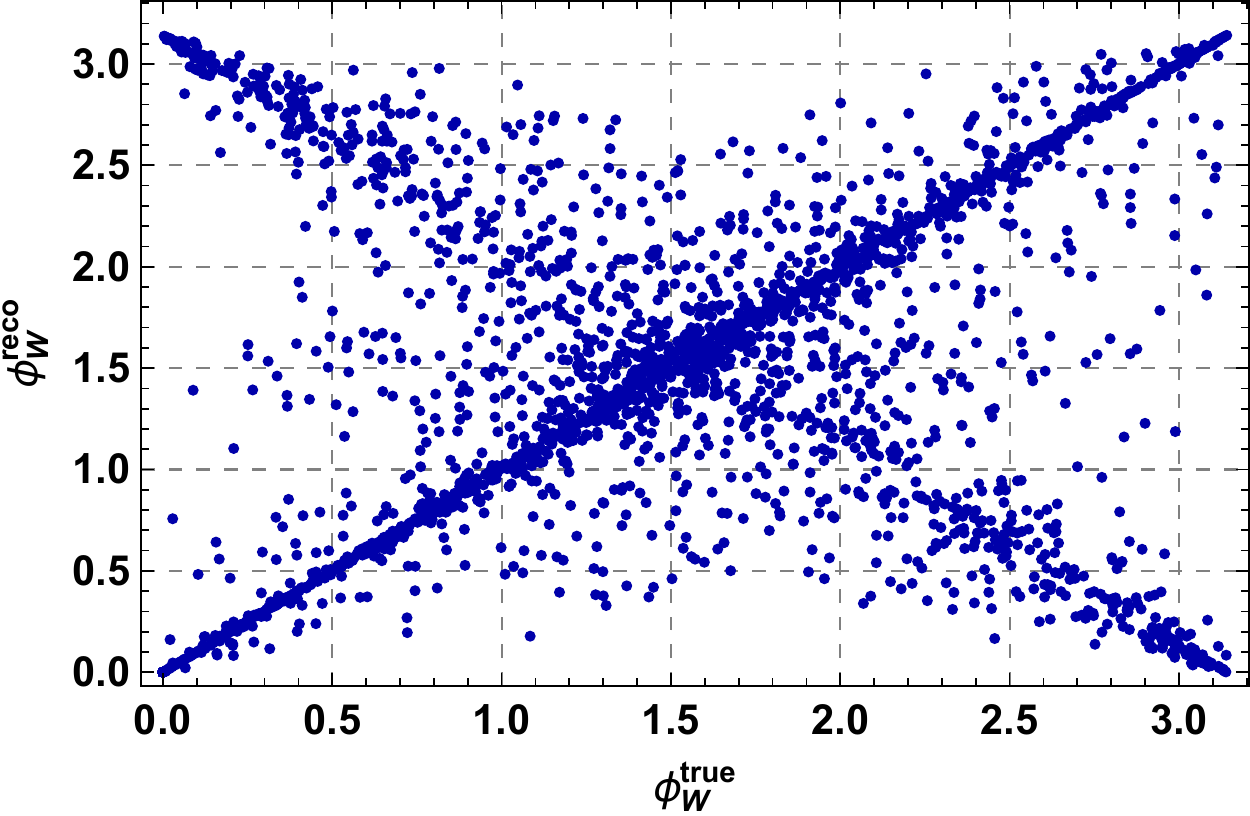}
\caption{Parton level values of the $\phi_W$ angle defined in Eq.~\eqref{eq:angledef} built assuming a randomly chosen $p_z^\nu$ solution against the same angle where the real, but experimentally unaccessible, value of $p_z^\nu$ has been used. The ambiguity $\phi_W\to \pi-\phi_W$ is manifest.
}
\label{fig:nu_ambig}
\end{figure}

Note that in the case of the $Z$ boson, since its coupling to left- and right-handed charged leptons are approximately equal, we cannot unanbiguosly  identify the helicities of the final state leptons. As a consequence   the normal vector $\hat n_{{\rm{decay}}}^Z$ is defined only up to an overall sign. By using the definition of  Eq.~\eqref{eq:angledef},  this translates into an ambiguity
\bea
\phi_Z\leftrightarrow \phi_Z-\pi.
\eea
None of the modulations of the Eq.~\eqref{eq:idealmodWZ} are however  affected by this  ambiguity, since they are functions of $2\phi_Z$.  Now let us look at the azimuthal angle of the 
leptons from the $W$ boson decay. Differently than for the $Z$ boson, in this case the helicities  of the final state leptons are fixed  by the pure left-handed nature of the EW interactions. 
 However in this case the azimuthal angle determination suffers from 
 a twofold ambiguity on the determination of the longitudinal momentum of the invisible neutrino, arising from the quadratic equation determining the on-shellness of the $W$ boson.  All together for boosted $W$ bosons this leads to the approximated ambiguity 
 \bea
\phi_W\to \pi-\phi_W
\label{eq:nu_ambig}.
\eea
This is illustrated in Fig.~\ref{fig:nu_ambig}, where we plot the $\phi_W$ angle constructed assuming a randomly chosen $p_z^\nu$ solution against the same angle where the real, but experimentally unaccessible, value of $p_z^\nu$ has been used. The ambiguity of Eq.~\eqref{eq:nu_ambig} clearly washes away the $\sin 2\phi_{W}$ modulations of Eq.~\eqref{eq:idealmodWA} and Eq.~\eqref{eq:idealmodWZ}.

Before concluding this Section a comment is in order. Our definition of the diboson 
scattering plane of Eq.~\eqref{eq:planes} strictly assumes a  $2\to 2$ scattering process, 
where the two vector bosons are produced back to back in the center of mass frame of the initial partons. In the case of real radiation 
emission, as in the case of the presence of initial state radiation jets, the diboson scattering 
plane has to be defined directly through the momenta of the two vector boson. However in 
the case of the $WZ$ and $W\gamma$ processes the determination of this plane will be 
again affected by the neutrino reconstruction ambiguities.  We then decide to use the
 definition of Eq.~\eqref{eq:planes} when building the azimuthal angles of  
Eq.~\eqref{eq:angledef} throughout our analysis. This is a good approximation, since only processes with a hard jet emissions, which are kinematically suppressed,
 can lead to a significant differences between the planes orientations.


\section{$pp\to W Z$ process}\label{sec:WZ}

We begin by studying the fully leptonic $pp\to W Z$ process at the LHC.  Before doing so we wish to describe to simulation environment which will also be used for the analysis of the fully leptonic $pp \to W\gamma$ process discussed in Sec.~\ref{sec:WA}.

\subsection{Details on the event simulation}

We simulate the hard scattering fully leptonic $pp\to W Z$ process via the  {\tt MadGraph5\_aMC\@NLO} platform~\cite{Alwall:2014hca} using the  {\tt HELatNLO\_UFO} model that have been implemented in the {\tt FeynRules} package~\cite{Alloul:2013bka} and exported under the {\tt UFO} format~\cite{Degrande:2011ua} by the authors of~\cite{Degrande:2016dqg}~\footnote{We thank the authors of~\cite{Degrande:2016dqg} for sharing their NLO model files with the addition of the CP-even and CP-odd operators of Eq.~\eqref{eq:operators} previous to the publication of their paper.}.
We perform our study at NLO in QCD commenting, when relevant, the differences with respect to the results obtained at leading-order as well as NLO with an extra jet radiation in the matrix-element (hereafter NLO+$j$), which partially mimics the next-to-next-to-leading-order (NNLO) accuracy. Parton showering and hadronization of partonic events has been performed with  {\tt PYTHIA8}~\cite{Sjostrand:2007gs}. Matching and merging between hard-scattering and parton shower have been performed through {\tt PYTHIA8} for the NLO case and  {\tt PYTHIA8} $+$ {\tt FxFx} algorithm as described in~\cite{Frederix:2012ps} for the NLO+$j$ case. We report in a compact way in Tab.~\ref{tab:tools} the summary of the tools used for each level of the perturbative expansions.  When analyzing the events, jets have been reconstructed via the anti-$\kappa_T$ algorithm~\cite{Cacciari:2008gp} with $\Delta R=0.4$ and a $p_T$ threshold of 20 GeV through the {\tt MadAnalysis5} package~\cite{Conte:2012fm} as implemented in {\tt MadGraph5\_aMC\@NLO}. 

While our event generation has been performed at the partonic level, we wish to  mimic (at least partially) detector smearing effects when building the angular variables used for our analysis, without performing a dedicated detector simulation for all our event samples. We do so as follows. We choose one event sample and compare, on an event by event basis, the values of the $\phi_Z$ and $\phi_W$ variables before and after having applied detector effects, which we have evaluated through the {\tt Delphes 3} package~\cite{deFavereau:2013fsa}. We build the distributions of the $\Delta \phi^{\rm smear}_{Z,W}=\phi_{Z,W}^{\rm parton}-\phi_{Z,W}^{\rm detector}$ difference and construct the corresponding probability distribution function. We approximate the latter with a three rectangles shape and dress the parton level values of the azimuthal angles with a  $\Delta \phi^{\rm smear}_{Z,W}$ evaluated with the computed probability. For concreteness we use the following functions
\footnote{We define the change of the angle due to the smearing to be in the interval $[-\frac{\pi}{2},\frac{\pi}{2}]$ due to the $2\phi$ periodicity of the modulation terms of Eq.~\eqref{eq:idealmodWZ} in the cross section.  }
\bea\label{eq:smearfwz}
&&\phi^{\rm smear}_Z= \phi_Z\pm \Delta\phi_Z^{\rm{smear}},~~~\Delta\phi_Z^{\rm{smear}}=
\left \{
\begin{array}{c}
 \left[0,0.2\right] \rm{with~ probability}~~ 0.68 \\
\left[0.2,\pi/2 \right ]\rm{with~ probability } ~~~0.32,
\end{array} 
 \right  . \nonumber\\
 && \nonumber\\
  && \nonumber\\
&&\phi^{\rm smear}_W= \phi_W\pm \Delta\phi_W^{\rm{smear}},~~~\Delta\phi_W^{\rm{smear}}=
\left \{
\begin{array}{c}
 \left[0,\pi/4\right] \rm{with~ probability}~~ 0.66 \\
\left[\pi/4,\pi/2 \right ]\rm{with~ probability } ~~~0.34 .
\end{array} 
 \right  .
\eea

\begin{table}
\begin{center}
\begin{tabular}{c|c|c|c}

Order  & Hard-scattering & Parton Shower & Jet Merging\\
    \hline\hline
LO 		&    \multirow{3}{*}{{\tt MadGraph5\_aMC\@NLO}}&\multirow{3}{*}{{\tt PYTHIA8}} & $\slash$ \\
NLO 		&    & & {\tt PYTHIA8}\\
NLO+$j$  	&    & & {\tt PYTHIA8}+{\tt FxFx}  \\

\end{tabular}
\caption{Summary of the tools used for the event generations at each order in the QCD perturbative expansion.}
\label{tab:tools}
\end{center}
\end{table}

\subsection{Comparison of perturbative expansions for the SM}\label{sec:SM_LO_NLO}

\begin{figure}[t!]
\centering
{\includegraphics[width=0.46\textwidth]{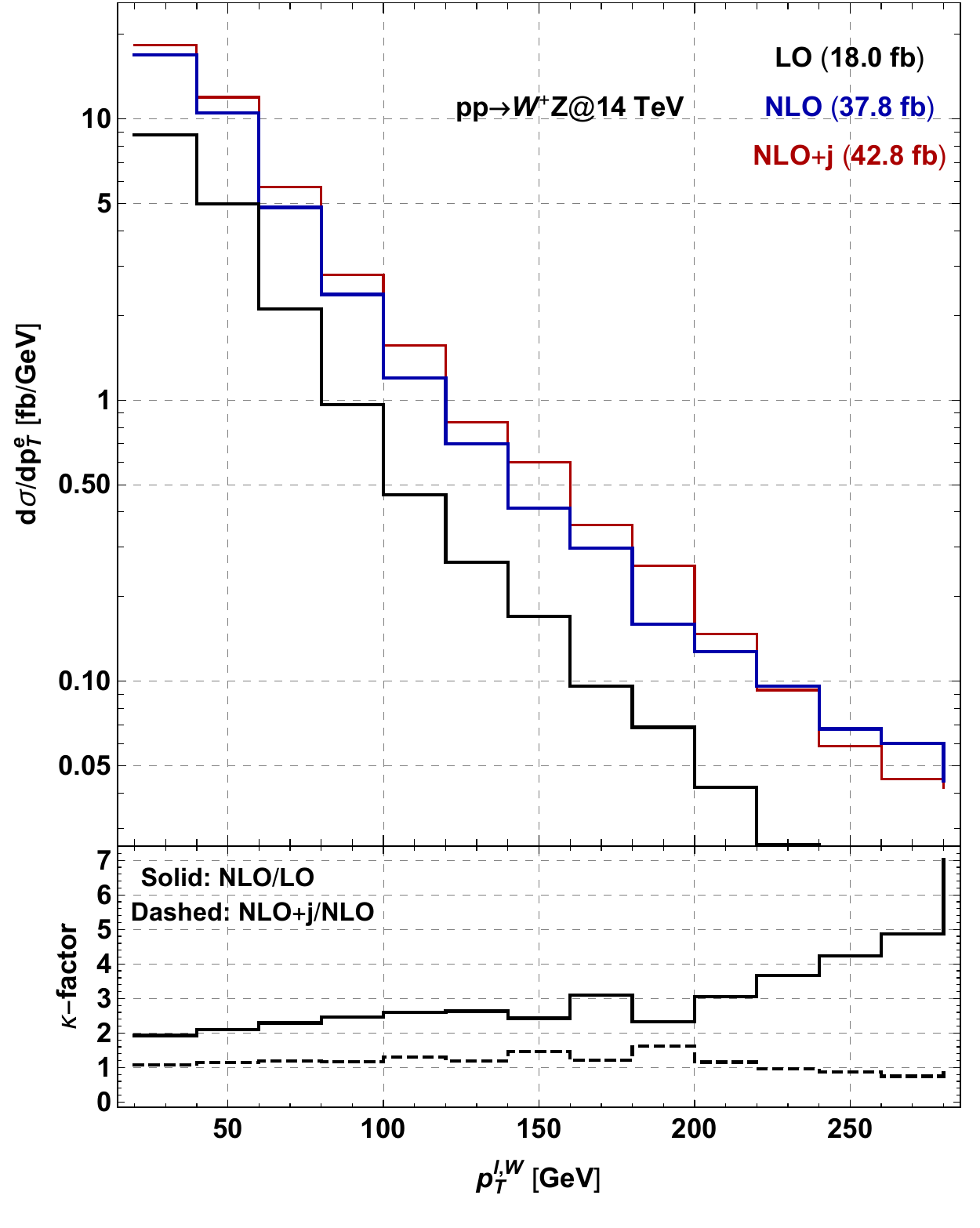}}\hfill
{\includegraphics[width=0.46\textwidth]{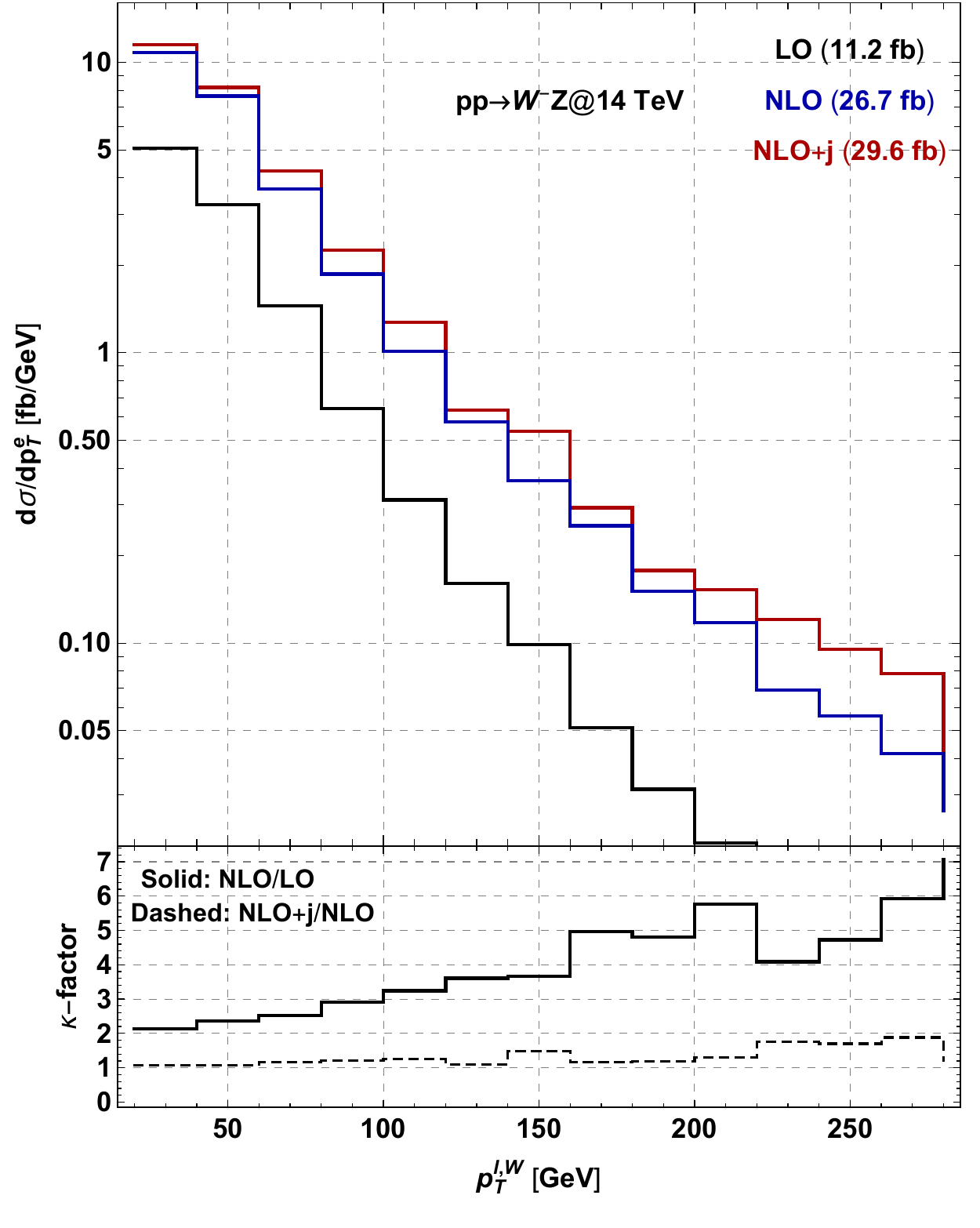}}
\caption{Differential distributions in function of the transverse momentum of the lepton arising from the $W^+$ (left) and $W^-$ (right) decay at the LO, NLO and NLO+$j$ accuracy for the SM fully leptonic  $pp \to W^\pm Z$ process. In the lower panels we show the NLO/LO and NLO+$j$/NLO differential cross section ratios.}
\label{fig:LO_vs_NLO_vs_NLOj}
\end{figure}

Before proceeding to the study of the sensitivity on the $O_{3W}$ and $O_{3\tilde W}$ operators we wish to validate our simulation framework against the existing literature for the case of the SM.
We consider the $p p \to WZ$ processes separately for the two $W$ boson charge signs at  LO, NLO and NLO+$j$ order for the LHC with a center of mass energy of 14 TeV. We force the $Z$ boson to decay into a muon pair and the $W$ boson into an electron and the associated neutrino. By applying only a 20 GeV cut on the transverse momenta of all visible leptons we obtain a cross section value at NLO and LO of $37.8\;$fb and $18.0\;$fb for the $W^+$ case and of $26.7\;$fb and $11.2\;$fb for the $W^-$ case. The addition of an extra jet in the matrix element increases these value of an extra $\sim 10\%$. These findings nicely agree with the latest results of~\cite{Grazzini:2017ckn}, computed for $\sqrt s=13\;$TeV. For the same processes we then compare the differential cross sections in function of the transverse momentum of the charged lepton from the $W$ decay, also reported in~\cite{Grazzini:2017ckn} for $\sqrt s=8\;$TeV.
By taking into account the parton luminosity rescaling factor between our and their center of mass energy (which is $\sim 2$ for the $q\bar q$ scattering of proton's valence quarks for $\sqrt{\hat s}=~300\;$GeV) we find an overall good agreement in the distributions shapes between our LO and NLO results and the ones of~\cite{Grazzini:2017ckn}, thus further validating our simulation framework. 
Again, we observe that there is a small difference between the NLO and NLO+$j$ calculations. Given the larger computation time needed for the latter simulation, we will present our results only at NLO accuracy in QCD commenting however, where relevant, what could be the effect of the extra real radiation on the processes under consideration.

\subsection{Sensitivity to the BSM operators}
\label{sec:WZ_res}

We now turn on the BSM operator $\mathcal{O}_{3W}$ and $\mathcal{O}_{3\tilde W}$ defined in Eq.~\eqref{eq:operators} and simulate LO and NLO events with the same strategy as for the SM case described in Sec.~\ref{sec:SM_LO_NLO}, and applying a final combinatorial factor to take into account all possibile final state flavor configurations involving the first two generation lepton flavor. 
 We generate events with only the CP-even or CP-odd operator different from zero, as well as events where both the operators are present, so as to determine the contribution to the cross section due to the interference of the two deformations. We closely follow the ATLAS experimental analysis of~\cite{ATLAS:2018ogj} and we define our signal region imposing the following sets of cuts:
 $p_T^{e}>20\;$GeV,  $p_T^{\mu}>15\;$GeV, $|\eta^{\mu,e}|<2.5$, $\Delta R(\ell,\ell)>0.2$, $\Delta R(\ell,j)>0.4$, where $\ell=e,\mu$ and the $p_T$ threshold for jets is $20\;$GeV. We further require that the same flavor opposite charge lepton pair reconstruct the $Z$ boson asking $|m_{\mu^+\mu^-}-m_Z|<20\;$GeV and we impose a cut of 30 GeV on the $W$ boson transverse mass~\footnote{
 The $W$ boson transverse mass is defined in the next Section in Eq.~\eqref{eq:MWT}.}. We then bin our events with respect to the $WZ$ system transverse mass, which we define as
\be\label{eq:mWZT}
(m_{{WZ}}^T)^2=\left(\sqrt{m_W^2+\sum_{i=x,y}(p_i^e+\slashed p_i)^2}+\sqrt{m_Z^2+\sum_{i=x,y}(p_i^{\mu^+}+ p_i^{\mu^-})^2}\right)^2 - \sum_{i=x,y}(p_i^e+\slashed p_i+ p_i^{\mu^+}+ p_i^{\mu^-})^2,
\ee where $\slashed p_i$ is the i-th component of the missing transverse momentum of the event.
We finally build the $\phi_Z$ and $\phi_W$ azimuthal angles as defined in Eq.~\eqref{eq:angledef} and 
categorize the events with respect to $\phi_Z$  and  $\phi_W$, both defined in the range 0 to $\pi$.
\begin{figure}[t!]
\centering
{\includegraphics[width=0.46\textwidth]{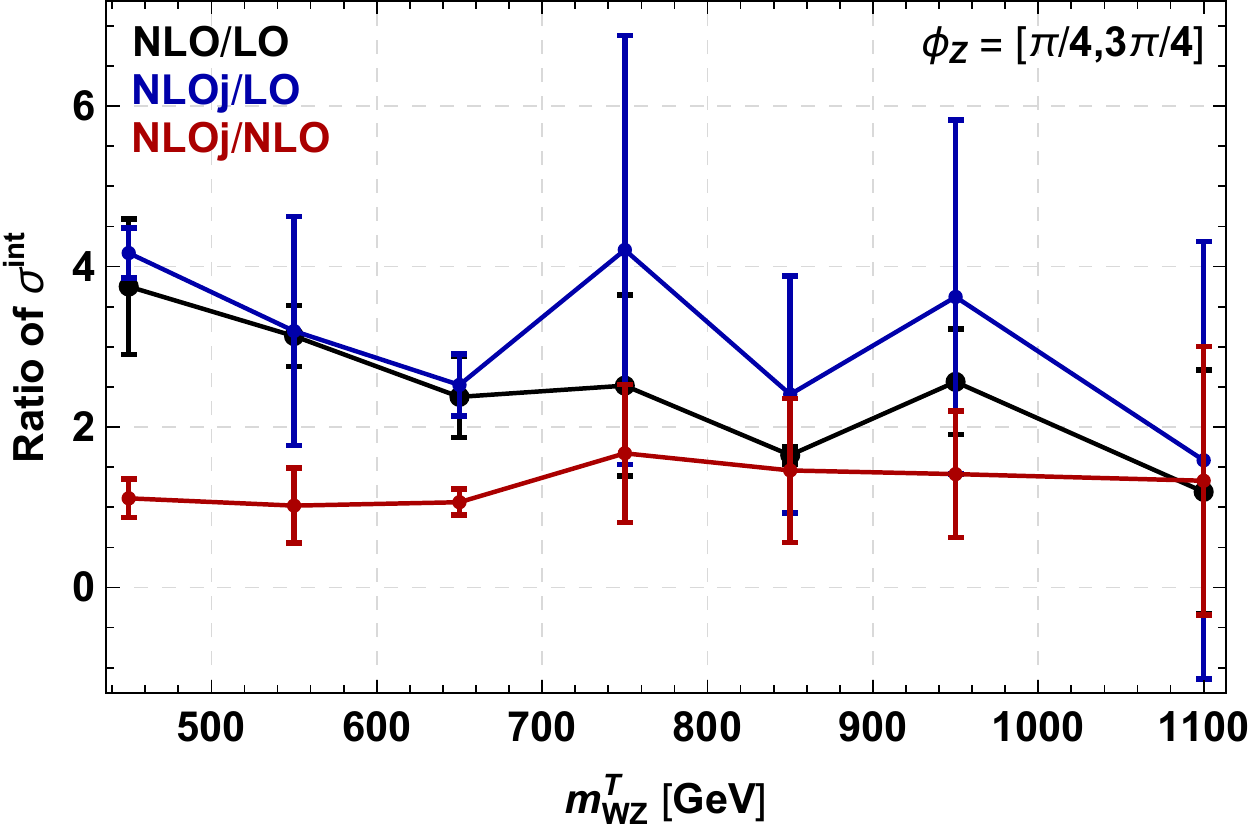}}\hfill
{\includegraphics[width=0.475\textwidth]{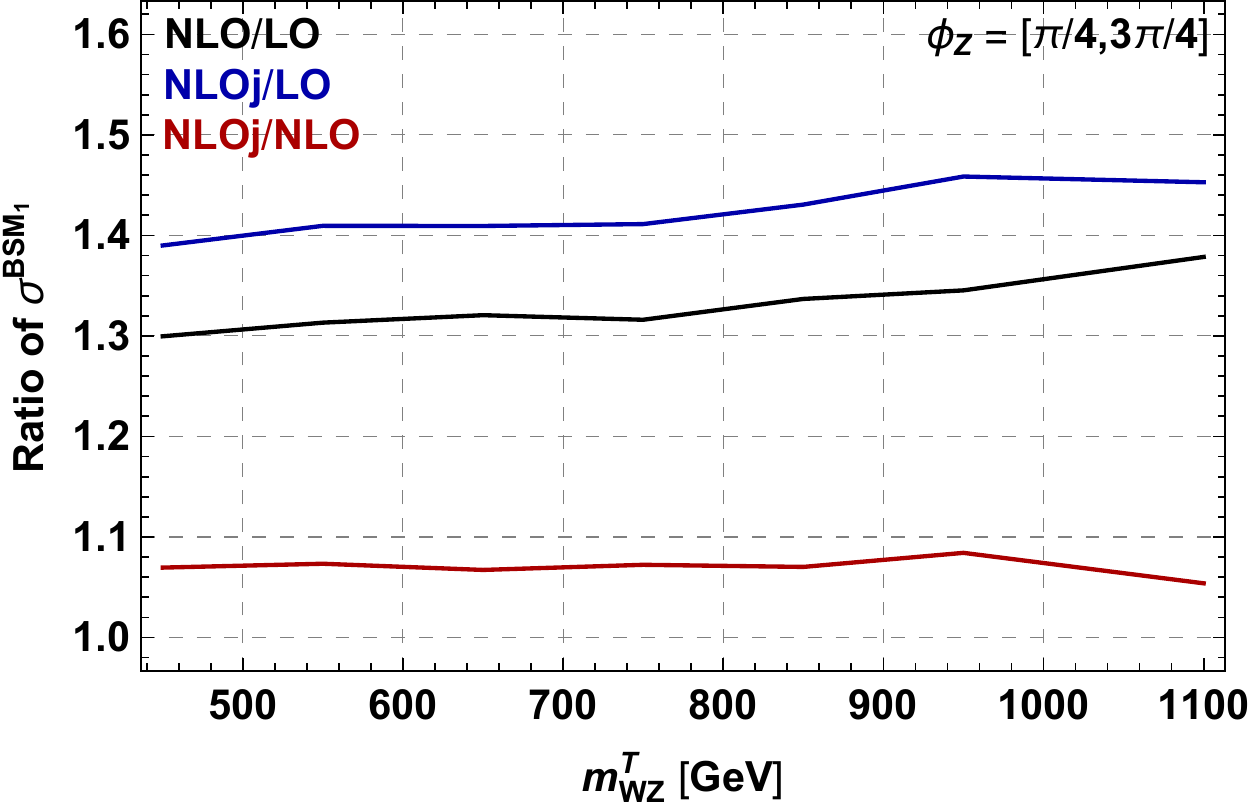}}
\caption{Differential distribution  for the NLO/LO, NLO+$j$/LO and NLO+$j$/NLO ratios of $\sigma^{\rm int}$ (left panel) and $\sigma^{\rm BSM_1}$ (right panel) in function of the $WZ$ transverse mass.}
\label{fig:LO_vs_NLO_vs_NLOj_a1a2}
\end{figure}

Now we can proceed to the analysis of the various BSM contributions. Generically the  production cross section in the presence of the operators  of Eq.~\eqref{eq:operators} is given by
\bea
\label{eq:xsecdef}
\sigma=\sigma_0+ \sigma^{\rm int} c_{3W}+\tilde \sigma^{\rm int}\tilde c_{3W}+ \sigma^{\rm BSM_1}c^2_{3W}+ \sigma^{\rm BSM_2}\tilde c^2_{3W}+\sigma^{\rm BSM_3} c_{3W} \tilde c_{3W}.
\eea
We firstly compare in Fig.~\ref{fig:LO_vs_NLO_vs_NLOj_a1a2}  the LO, NLO and NLO+$j$ interference, $\sigma^{\rm int}$, (left) and quadratic, $\sigma^{\rm BSM_1}$, (right) terms of the cross section in presence of the CP-even operator $\mathcal{O}_{3W}$ in the angular region $\phi_Z \in [\frac{\pi}{4},\frac{3\pi}{4}]$ in function of the $m_{WZ}^T$. 
We observe that for the pure BSM term  the $\kappa$-factor between NLO and LO is $\sim 1.3$, only mildly growing with the partonic energy of the process, and that the addition of an extra jet in the matrix element only provide a small increase, around 5\%, with respect to the NLO process, similarly to what has been found for the inclusive process in the SM case.
On the other side, for the interference case, the $\kappa$-factor shows a slightly decreasing pattern with the energy of the system, reaching a value of $\sim 2$ for $m_{WZ}^T\sim 1\;$TeV.
Furthermore helicity selection rules are not applicable at NLO level leading to a mild restoration of the interference effects between the SM and BSM contributions \cite{Azatov:2017kzw,Chiesa:2018lcs}.  
Additionally  the off-shellness of the vector bosons also leads to the restoration of the interference, with the strength of the effect scaling as $g^2$~\cite{Helset:2017mlf}, similarly to the effect of the one loop electroweak corrections, which we ignore in the present study.
We can notice that the statistical error in the determination of the NLO$+j$/LO and NLO$+j$/NLO ratios for $\sigma^{\rm int}$ can be quite large, almost around $50\%$, due to bigger uncertainties in the analysis of the interference at NLO$+j$ accuracy. However, these statistical fluctuations do not affect the precision on the results we will show for the $c_{3W}$ and $\tilde c_{3W}$ bounds: they are obtained at NLO, without an extra jet emission, and at such level the uncertainty on the interference is smaller, around $10\%$.

\begin{figure}[t!]
\centering
{\includegraphics[width=0.46\textwidth]{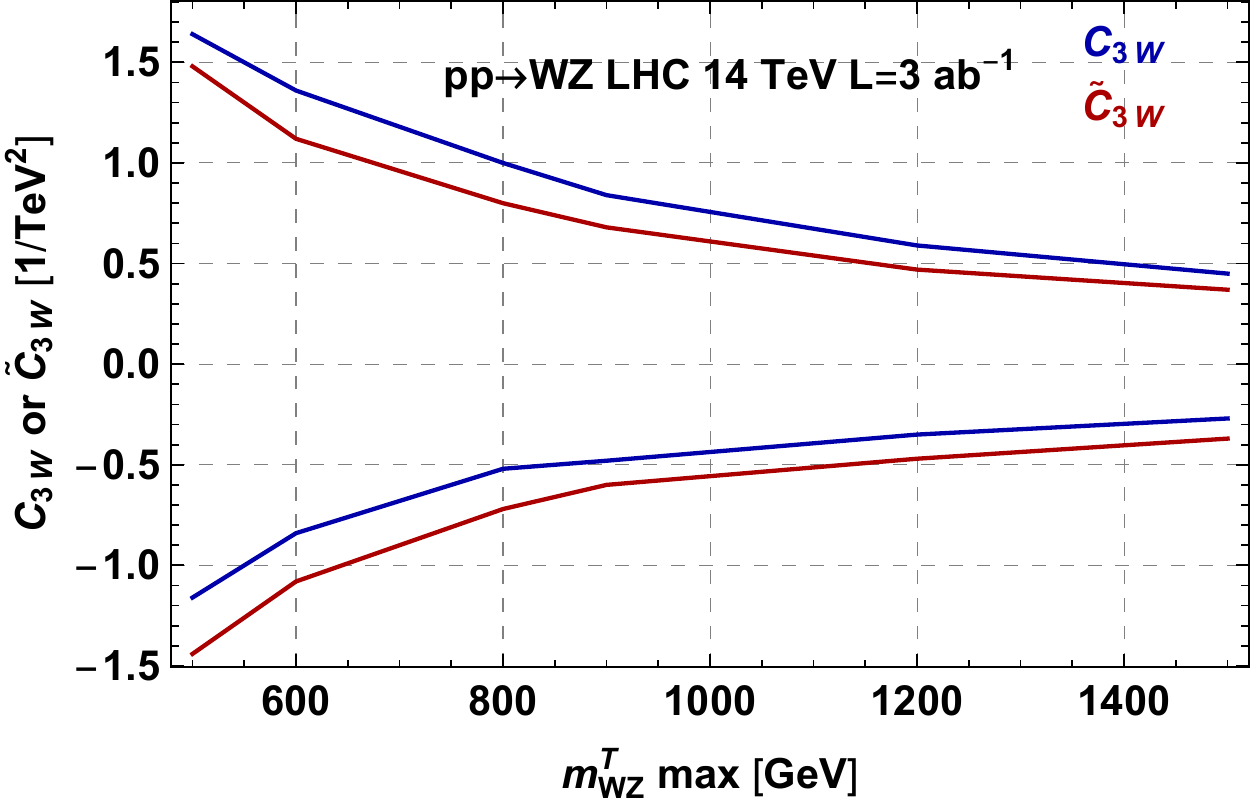}}\hfill
\caption{95\%  bound on the $c_{3W}$ and $\tilde c_{3 W}$ Wilson coefficients computed with four and two equally spaced angular bins for $\phi_Z$ and $\phi_W$ respectively, in function of the largest $WZ$ system transverse mass bin used for the 14 TeV LHC with 3000 fb$^{-1}$ of integrated luminosity.  A systematic error of 5\% has been assumed.  }
\label{fig:WZ_bound}
\end{figure}

We now proceed in setting the bounds on the $c_{3W}$ and $\tilde c_{3 W}$ Wilson coefficients as follows. We categorize our events with respect to four angular $\phi_Z$ and two $\phi_W$ bins, equally spaced in the range 0 to $\pi$, and  with respect to the $WZ$ system transverse mass, with $m_{WZ}^T$ bins between [0,1000] GeV in steps of 100 GeV, [1000,1200] GeV and [1200,1500] GeV.
We consider only the SM irreducible $WZ$ background, which is the main source of background for this process~\cite{ATLAS:2018ogj}, and we impose a global efficiency of 0.6 for reconstructing the final state for all lepton flavor combinations.
Then, by assuming a Poissonian distributed statistics,  we perform a Bayesian statistical analysis  estimating the systematical  error through one nuisance parameter (see~\cite{Azatov:2017kzw} for more details). We find that the binning in $\phi_W$ has a marginal impact on the limits determination, which is due to the large smearing on the $\phi_W$ variable with respect to $Z$ decay products azimuthal angles. 
Binning our events with respect to $\phi_Z$, $\phi_W$ and $m_{WZ}^T$, we obtain the 95\% posterior probability limits~\footnote{These limits are obtained by marginalizing on the value of the other Wilson coefficient.} on  
$c_{3W}$ and $\tilde c_{3 W}$ shown in Fig.~\ref{fig:WZ_bound}. The limits are shown in function of the maximum $m_{WZ}^T$ bin value used for the computation of the bounds and for an integrated luminosity of 3000~fb$^{-1}$, {\emph{i.e.}} at the end of the high luminosity phase of the LHC, assuming a systematic error of 5\%.

\begin{figure}[t!]
\centering
\includegraphics[scale=0.55]{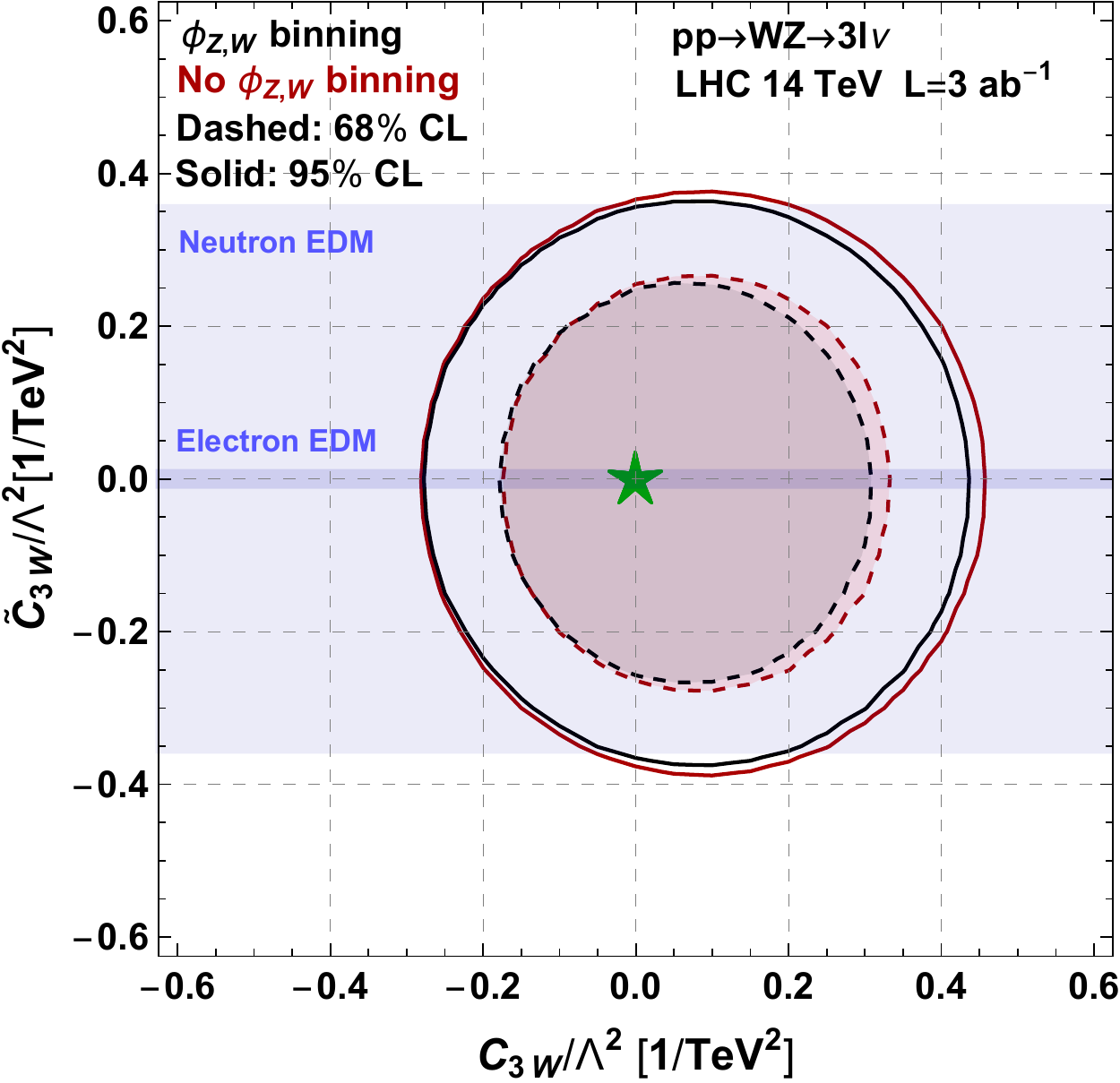}\hfill
\includegraphics[scale=0.55]{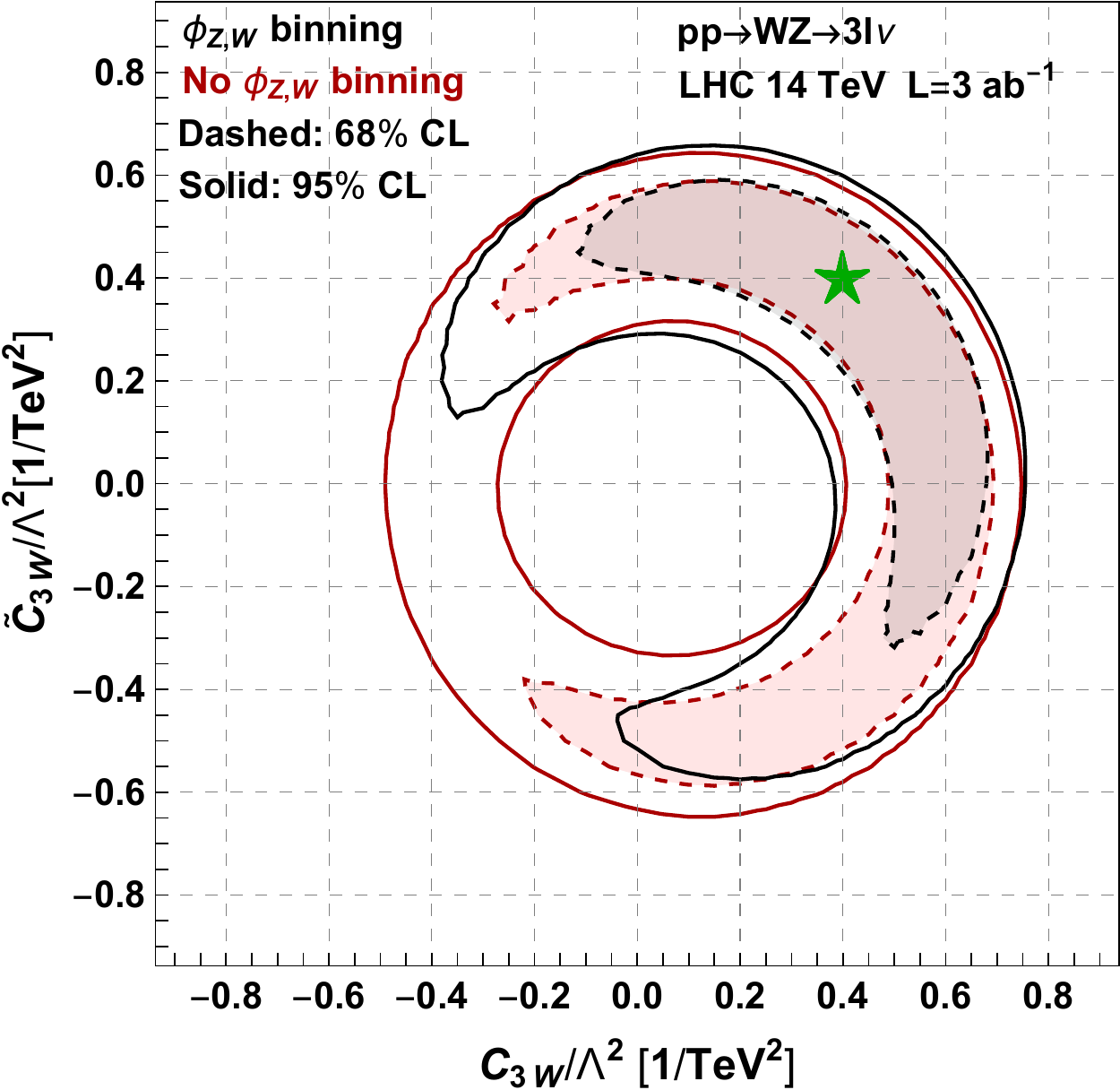}\\
\includegraphics[scale=0.55]{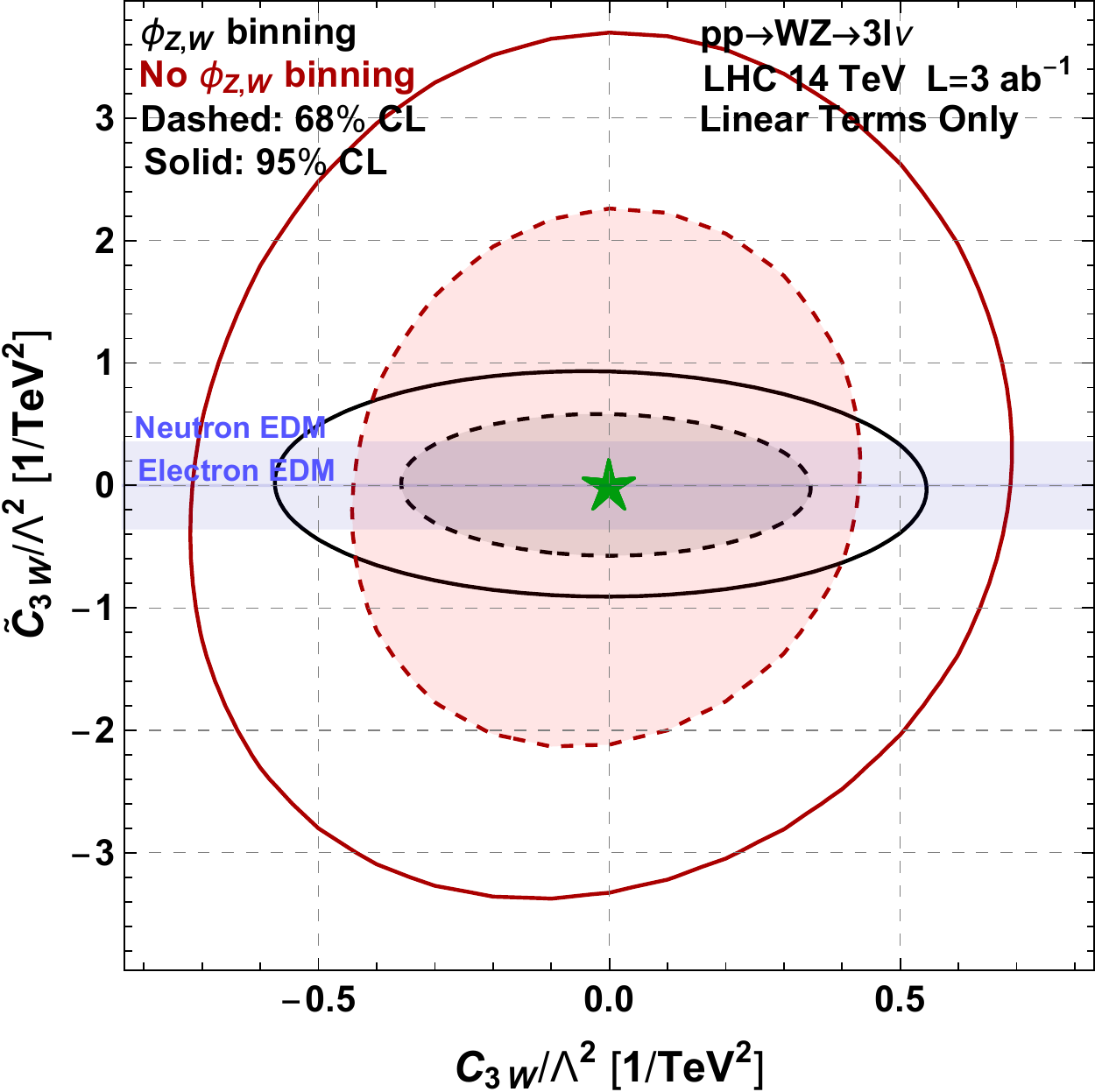}
\caption{ 68\%  (dashed) and 95\%  (solid) posterior probability contours for the analysis with (black) and without (red) the binning in the $\phi_Z$ and $\phi_W$ angles, see main text for more details.
The left and right hand upper plots are obtained assuming the SM and a BSM signal with $c_{3W}=\tilde c_{3W}=0.4$, both represented by a green star.
The light and dark shaded blue correspond to the limits obtained by the non observation of a neutron and electron EDM discussed in Sec.~\ref{sec:EDM}.
On the lower plot for illustration purposes we present the exclusion contours assuming only the linear terms in the EFT expansion.
 Only events with $m^T_{WZ} <1.5 $ TeV are used.}
\label{fig:WZ_ellipses}
\end{figure}

We then fix a maximum value of 1500 GeV for the $m_{WZ}^T$ bin considered and we show in Fig.~\ref{fig:WZ_ellipses} the 68\% and 95\%  limits in the $c_{3W}-\tilde c_{3 W}$ plane assuming the SM (left panel) or a signal injection with $c_{3W}=\tilde c_{3 W}=0.4\;$TeV$^{-2}$, again with a systematic uncertainty of 5\% and an integrated luminosity of 3000 fb$^{-1}$.
There the black and red curves correspond to the probability contours with and without the binning in the $\phi_Z$ and $\phi_W$ angles and the shaded areas in the left panel correspond to the bounds derived from the non observation of a neutron (dark blue) and electron (light blu) EDM, discussed in Sec.~\ref{sec:EDM}.

We observe that the use of the azimuthal variables marginally improves on the limits when the SM is assumed.  This comes out from the combination of three different effects. Firstly, we are considering both the linear and the quadratic term in the EFT expansion, where the latter is not affected from the helicity selection rules cancellation. Secondly the helicity selection rules are violated by QCD NLO effects. Lastly, the imposition of kinematic cuts to select the analysis signal region have also the 
effect of restoring the interference between the SM and the BSM amplitude. 
Indeed, we have checked that  some of the cuts lead to a partial selection of the azimuthal 
angles. We postpone 
the discussion of this effect to   Sec.~\ref{sec:mod_cuts} when we discuss the $W\gamma$ process, since the effect is much stronger and
 the smaller number of final state particles makes it easier to understand the kinematic origin of this behavior. 
We notice however that the use of the azimuthal angles is crucial in the case of a signal discovery at the LHC.
As illustrated in the right panel of Fig.~\ref{fig:WZ_ellipses} this variable can in fact be used to disentangle the contribution of the $O_{3W}$ and $O_{3\tilde W}$ operators as well as to measure the sign of the Wilson coefficients.

At last we would like to comment on the importance of the linear terms in the the expansion of the cross section in Eq.~\eqref{eq:xsecdef}. We can see that the binning in the azimuthal angles increases the sensitivity on the $O_{3\tilde W}$ by a factor  $\sim 4$, while it has a marginal improvement on $ O_{3W}$, due the modulation from cuts effect discussed in the Sec.~\ref{sec:mod_cuts}. Comparing the "linear" and "quadratic" bounds we can see that the former are roughly factor of two worse for both the $O_{3 W}$ and the  $O_{3\tilde W}$ operators. This means that our analysis  can be applied only to the UV completions where the  contribution of the dimension eight operators  is smaller than both the quadratic and linear dimension six terms. Anticipating the results of the Sec.~\ref{sec:WA} and Sec.~\ref{sec:LHC27},  we find that for  $W\gamma$ analysis at 14 and 27 TeV and for $WZ$ analysis at 27 TeV the bounds are dominated by the linear terms.


\section{$pp\to W^\pm \gamma$ process}\label{sec:WA}

We next turn to another process which can be used to test the  CP-Even and CP-Odd operators of Eq.~\eqref{eq:operators}: $pp\to W^\pm \gamma$. As for the $WZ$ case, also here we consider a fully leptonic  final state which, despite having a smaller branching ratio and the presence of an invisible neutrino, is generally a cleaner channel with respect to the hadronic counterpart. Having validated our simulation framework for the $WZ$ case, we do not perform a comparison of the LO and higher orders samples for the $W\gamma$ process, and we consider from the beginning of our discussion the event samples generated at NLO accuracy.

\subsection{Modulation from cuts}\label{sec:mod_cuts}

\begin{figure}[t!]
\centering
\includegraphics[scale=0.55]{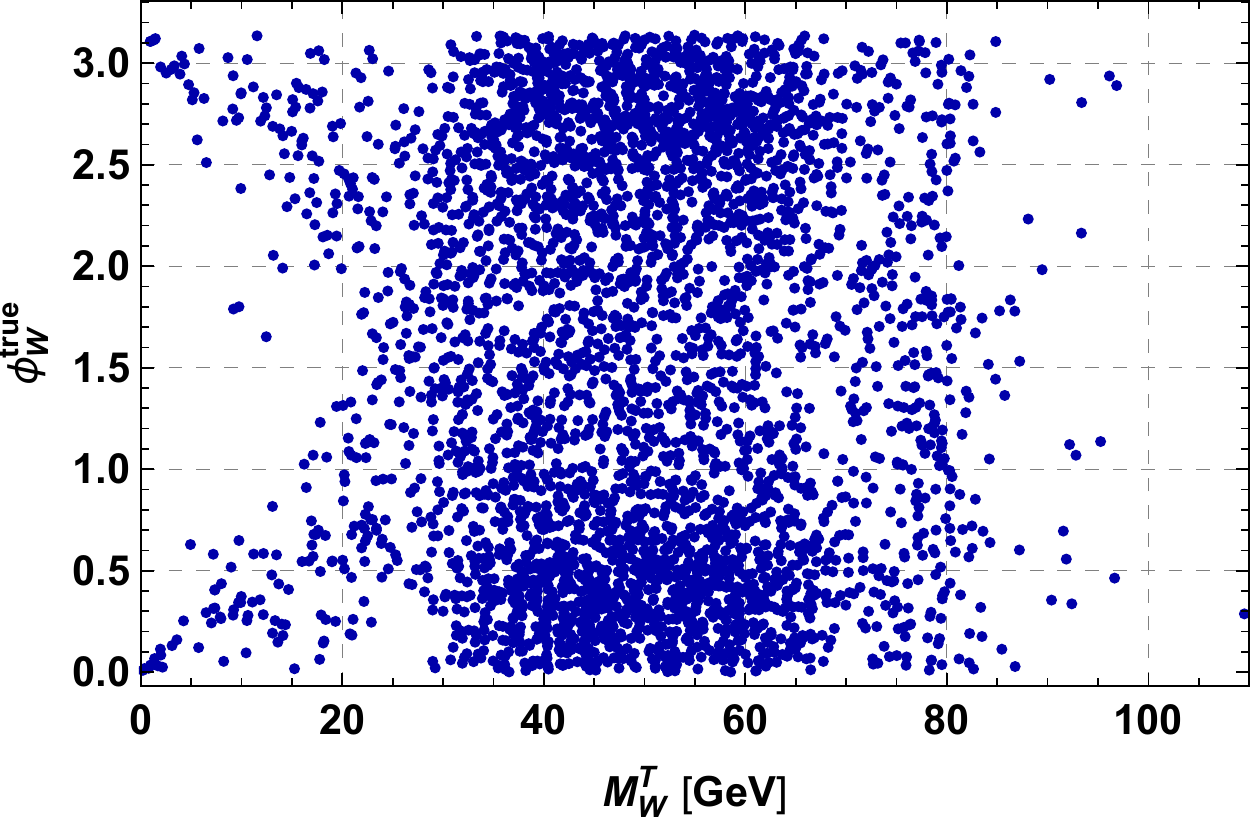}\hfill
\includegraphics[scale=0.55]{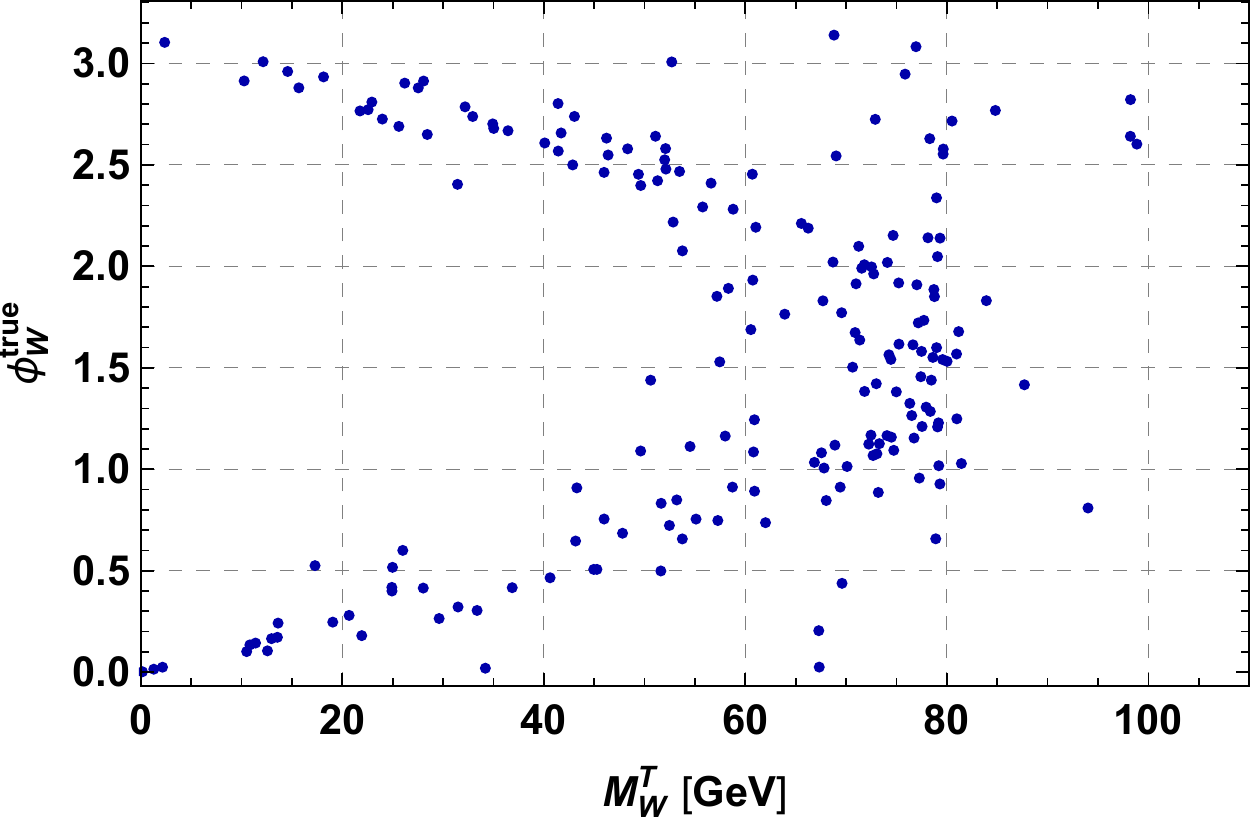}
\caption{Distribution of the azimuthal angle $\phi_W$ vs transverse mass of the W bososn $M_W^T$. Left - no other cuts are imposed, right additional cut on the $p_T^\gamma>100$ GeV  is required. \label{fig:modcut}}
\end{figure}

Before proceeding with the analysis, we comment here on the partial restoration of the interference between the SM and the BSM amplitudes arising from the imposition of certain kinematic cuts, which we anticipated in Sec.~\ref{sec:WZ_res}.
 Let's consider for example the cut on the $W$ boson transverse mass which is imposed in the experimental analysis~\cite{Chatrchyan:2013fya} and which is defined as
 \bea
\label{eq:MWT}
(M^T_W)^2=(p_T^e+\slashed p_T)^2-(\vec p_T^e+\slashed{\vec{p}}_T)^2
\eea
where $\slashed{\vec{p}}_T \approx \vec{p}_T^\nu$. By looking at the dependence of the azimuthal angle $\phi_W$ with respect to the transverse mass $M^T_W$ illustrated in the two panels of Fig.~\ref{fig:modcut} we observe that there is a strong correlation between the two variables. In the left panel all events within the detector kinematic acceptance are shown, while in the right panel we additionally impose $p_T^\gamma>100\;$GeV. 
In both plots we see that a small  $M^T_W$ is in correspondence with a value of 0 or $\pi$ for $\phi_W$. On the other side, for large $p_T^\gamma$, a cut on the $W$ boson transverse mass automatically selects events in the azimuthal bin $[\pi/4, 3\pi/4]$.
These two behaviors can easily be understood analytically.

Let's consider first the $M_W^T \sim 0$ case. In this limit the 
transverse momenta of the decay products of the $W$ boson are parallel:
\bea
\label{eq:limit_mw0}
M^W_T=0 \quad \Rightarrow \quad  \vec p_T^e  \parallel \vec p_T^\nu \parallel \hat  a
\eea
where $\hat a$ is a unit vector in the transverse plane. The momenta of the  $W$ boson and the charged lepton can be decomposed in a transverse and longitudinal part as 
\bea
&& \vec p_W= \alpha_W \hat a+\beta_W \hat z\nonumber\\
&& \vec p_e= \alpha_e \hat a+\beta_e \hat z
\eea
where $\hat z$ is a unit vector parallel to the beam line and $\alpha_{W,e}$ and $\beta_{W,e}$ are two real coefficients. Then Eq.~\eqref{eq:limit_mw0} fixes  the normals to the scattering plane and the decay planes, see Eq.~\eqref{eq:planes}, to be parallel
\bea
&& \vec n_{{\rm{decay}}}\propto \vec p_\nu\times \vec p_e \propto\vec p_W \times \vec p_e\parallel \hat a \times \hat z\\
&& \vec n_{{\rm{scat.}}} \propto \vec p_W \times \hat z \parallel \hat a\times z\nonumber
\eea
so that the azimuthal angle can only take the values of 0 or $\pi$.
In the high energy regime we can also understand the correlation shown in  the right panel of Fig.~\ref{fig:modcut} in the $M^W_T\sim M_W$ limit. Indeed let us assume that the $W$ boson is strictly on shell. Then the condition $M^W_T= M_W$ leads to
\bea
\label{eq:mtmw}
\frac{|\vec p_T^e|}{|\vec p_T^\nu|}=-\frac{p_z^e}{p_z^\nu}.
\eea
 Let us  consider the limit $p_T^W\gg p^W_z$, which is equivalent to requiring $p_T^\gamma\gg p^\gamma_z$. This limit in combination with the condition in Eq.~\eqref{eq:mtmw}
forces $p_T^{e,\nu}\gg p_z^{e,\nu}$. Hence in this case the normal to the decay plane will be always along the $\hat  z$ direction, so that the  azimuthal angle will take a value equal to $\pi/2$. 
All together we see that a high $M_T^W$ cut, together with the requirement of a large photon transverse momentum, lead to the automatic selection of a preferred azimuthal angle bin. 
In the analysis that we describe in the next Section we will bin the events in function of the transverse mass of the $W\gamma$ system, for analogy with what has been done for the $WZ$ case, where we have used the $m_{WZ}^T$ variable of Eq.~\eqref{eq:mWZT}. However for a $2\to 2$ scattering there is a one to one correlation between the $W$ boson and the photon transverse momenta. Hence, by selecting bins with high $M_{W\gamma}^T$ we automatically select events with high $p_T^\gamma$ which, as shown above, lead to the selection of events where $\phi_W \sim \pi/2$.

It is important to stress that a cut on the $W$ boson transverse mass that we have discussed is imposed in the experimental analysis that we 
consider~\cite{Chatrchyan:2013fya}. This kinematic selection is used to suppress 
backgrounds arising from processes without genuine missing transverse momentum, such 
as the overwhelming QCD $\gamma j$ background where a jet is misidentified as a lepton.
 Hence this {\emph{modulation from cuts}} behavior is always present when performing a real experimental analysis.
This is  an important effect which has been overlooked in similar studies in the previous literature and that leads to an enhanced sensitivity with respect to what is naively expected. A similar effect also occurs  in  $WZ$ channel process discussed in Sec.~\ref{sec:WZ_res}  and the plots on Fig.~\ref{fig:WZ_ellipses} reflect this property. However quantitatively we find it to be less important  than in the $W\gamma$ case.

\subsection{Sensitivity to the BSM operators}\label{sec:WA_res}

We now proceed to the analysis of the $W\gamma$ final state closely following the 7 TeV CMS results reported in~\cite{Chatrchyan:2013fya}, where a measurement of the $W\gamma$ inclusive cross section has been performed.
 As a first step we generate fully leptonic $W\gamma$ events for a center of mass energy of 7 TeV and we apply the same cuts enforced in the considered CMS search.
In particular CMS required the presence  of a lepton with $p_T>35\;$GeV and $|\eta|<2.5$ and of a photon with $p_T>15\;$GeV and $|\eta|<2.1$ and asked for a separation $\Delta R(\ell,\gamma)>0.7$. A cut on $M_T^W>70$\;GeV is also applied that, as mentioned, strongly suppresses the backgrounds from processes without genuine missing transverse energy. 
Then by comparing our NLO predictions with the results of~\cite{Chatrchyan:2013fya} we extract the efficiencies for reconstructing the $\ell \gamma$ final state, which we quantify to be 
0.45 for the electron and 0.7 for the muon.
 We then use the same efficiency values for the case of the 14 TeV LHC~\footnote{We have 
imposed in this case a 20 GeV cut $p_T^\gamma$ at generator level.}.
In order to estimate the detector effects on the determination of the azimuthal angle we follow exactly the same procedure as for the $WZ$ process (see Eq. \ref{eq:smearfwz}) and we find the the following smearing function
\bea
\phi^{\rm{smear}}_{W}= \phi_W\pm \Delta\phi_W^{\rm smear},~~~\Delta\phi^{\rm smear}_W=
\left \{
\begin{array}{c}
 \left[0,0.4\right] \rm{with~ probability}~~ 0.63 \\
\left[0.4,\pi/2 \right] \rm{with~ probability }~~~0.37.
\end{array} 
 \right  . 
\eea

\begin{figure}[t!]
\centering
{\includegraphics[width=0.46\textwidth]{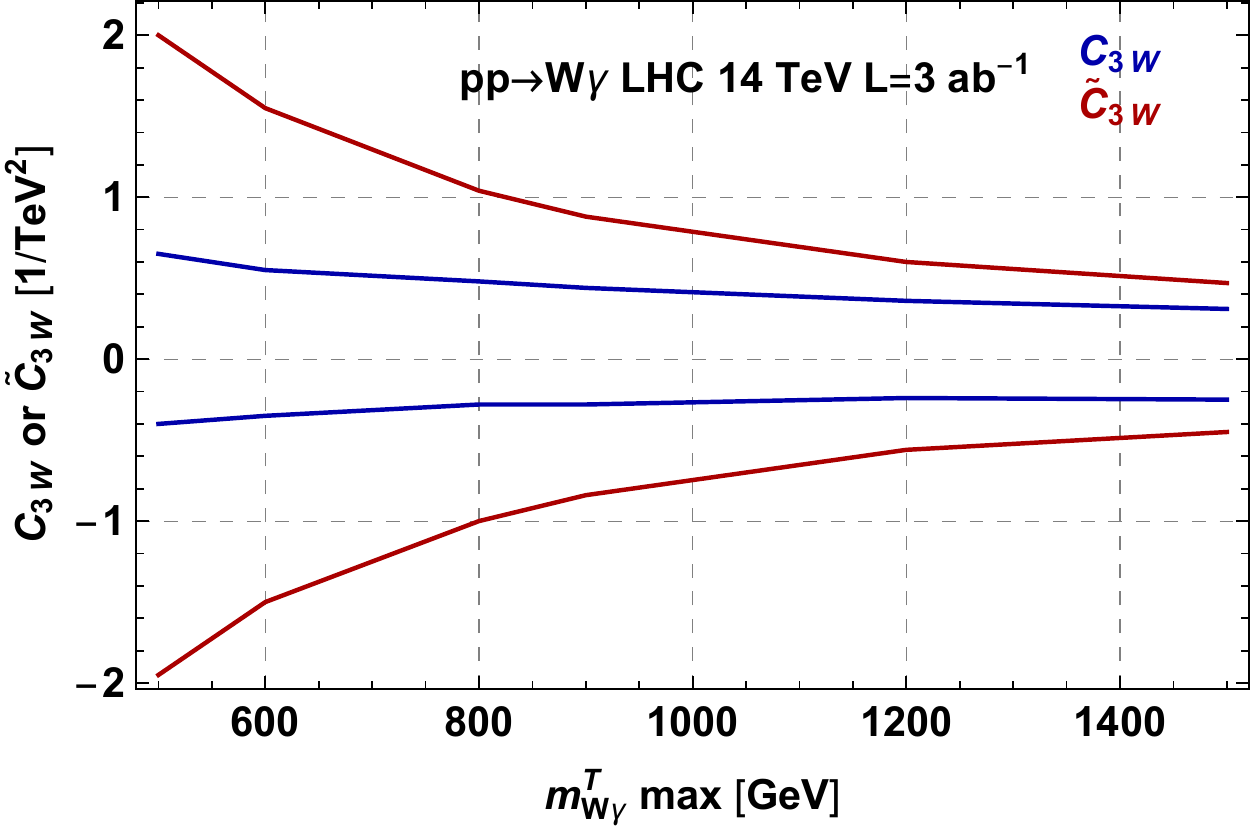}}\hfill
\caption{
95\%  bound on the $c_{3W}$ and $\tilde c_{3 W}$ Wilson coefficients computed with angular $\phi_W$ bins  (defined in the text) in function of the largest $W\gamma$ system transverse mass bin used for the 14 TeV LHC with 3000 fb$^{-1}$ of integrated luminosity.  A systematic error of 5\% has been assumed.}
\label{fig:WA_bound}
\end{figure}

\begin{figure}[t!]
\centering
\includegraphics[scale=0.55]{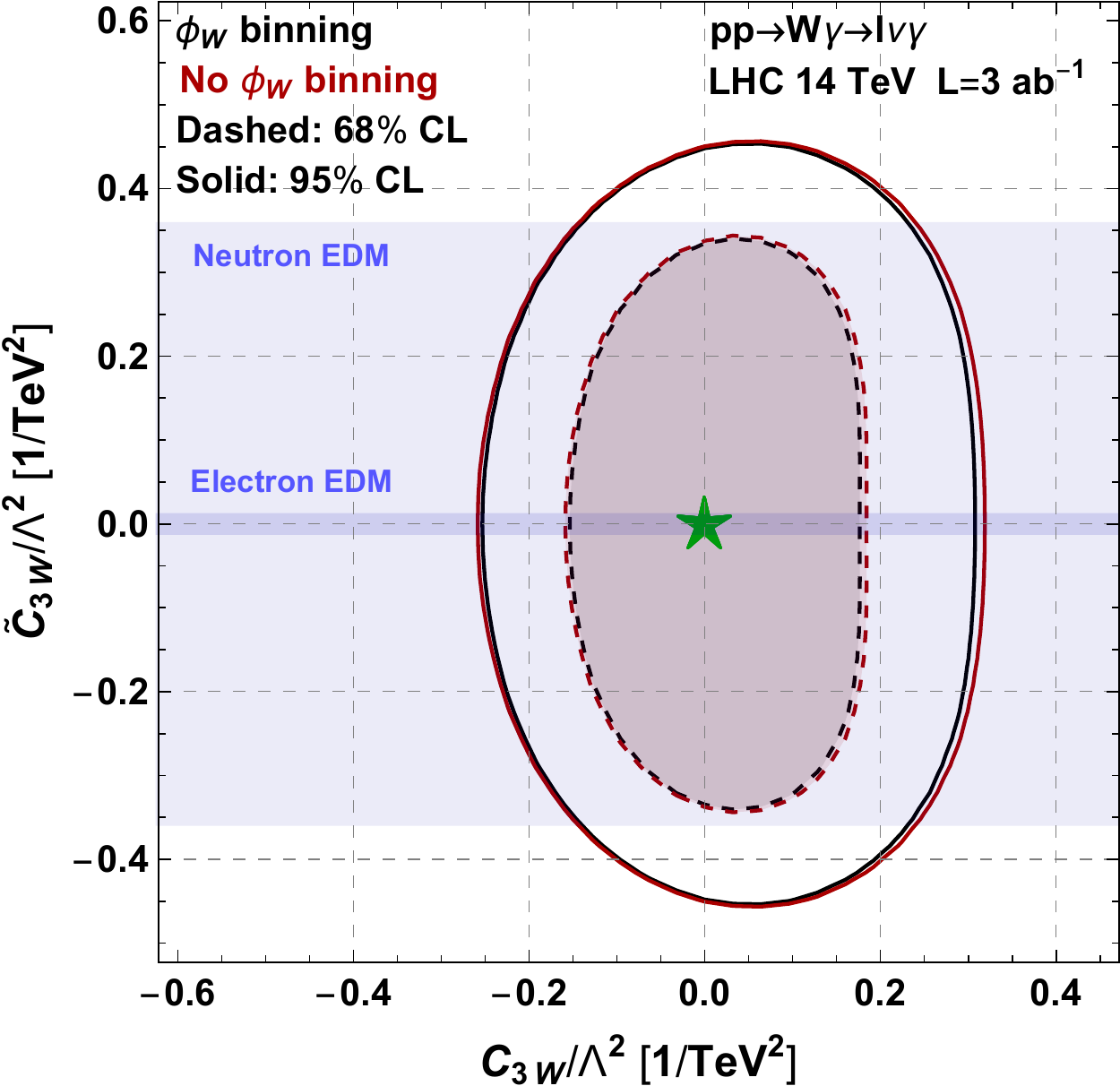}\hfill
\includegraphics[scale=0.55]{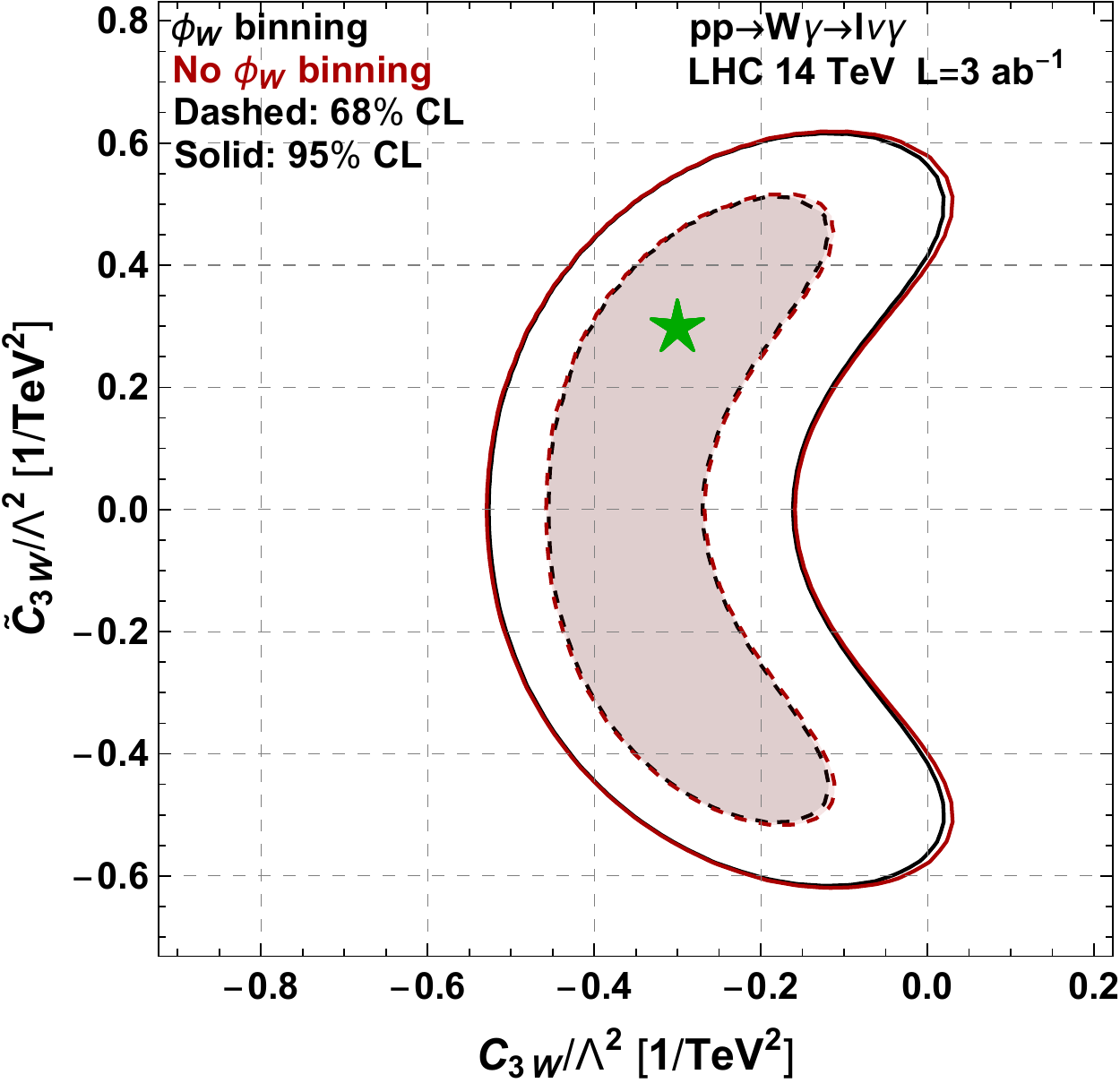}\\
\includegraphics[scale=0.55]{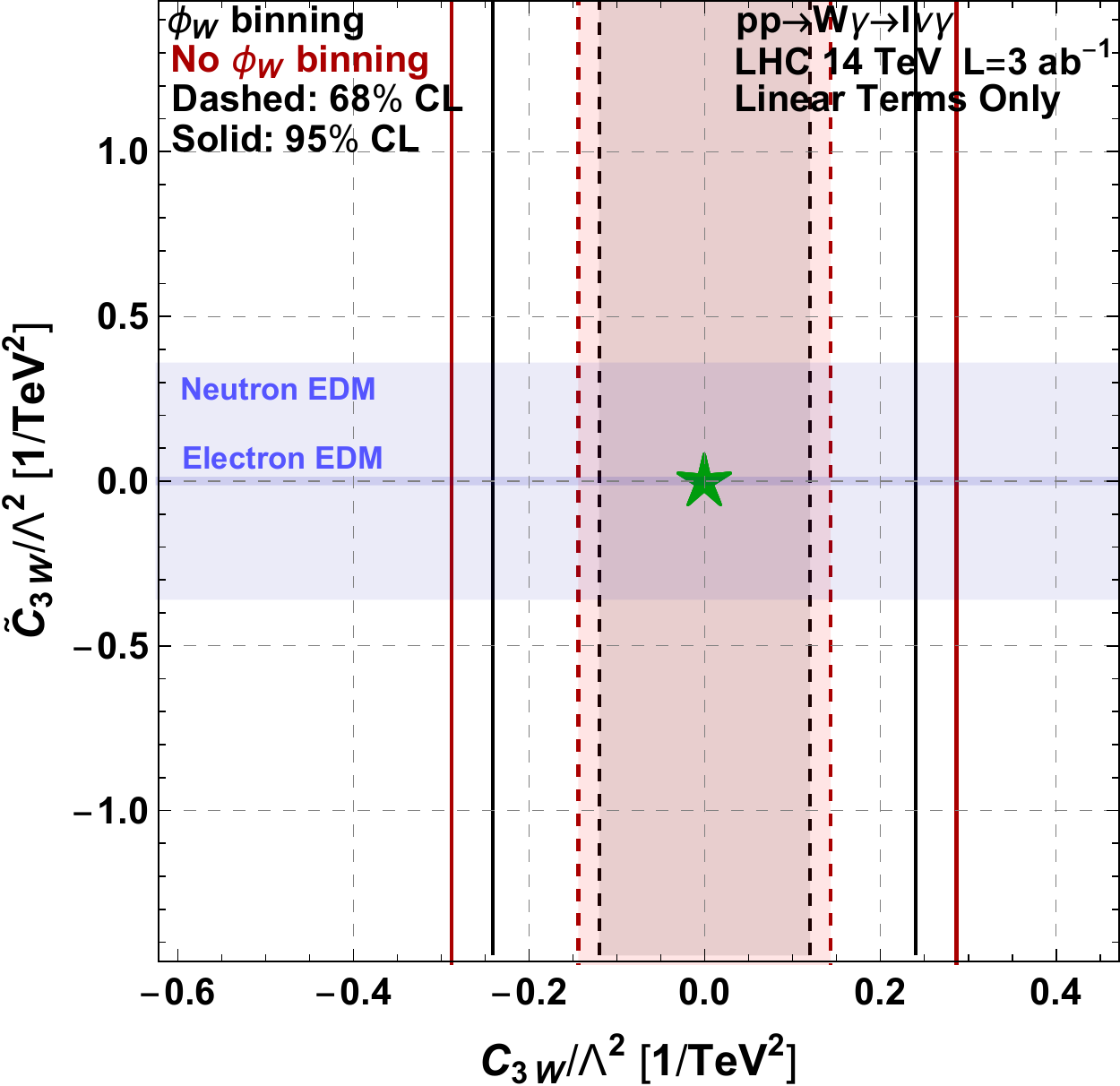}
\caption{ 68\%  (dashed) and 95\%  (solid) posterior probability contours for the analysis with (black) and without (red) the binning in then $\phi_W$ angle, see main text for more details.
The left and right hand upper plots are obtained assuming the SM and a BSM signal with $-c_{3W}= \tilde c_{3W}=0.3$, both represented by a green star. The light and dark shaded blue correspond to the limits obtained by the non observation of a neutron and electron EDM discussed in Sec.~\ref{sec:EDM}.
On the lower plot for illustration purposes we present the exclusion contours assuming only the linear terms in the EFT expansion.
Only events with $M_T^{W\gamma} <1.5 $ TeV are used.}
\label{fig:WA_ellipses}
\end{figure}

We notice that in the case of the $W\gamma$ process the irreducible SM background
makes only $\sim 50\%$ of the total event rate~\cite{ Chatrchyan:2013fya}. For this reason in our analysis we consider an equal yield for the irreducible and reducible background~\footnote{
This has been practically done by multiplying by a factor of two the $\sigma_0$ coefficients of the  Eq.~\eqref{eq:xsecdef} without touching the interference terms.
}. Clearly the reducible background does not interfere with the BSM operators under study, while the irreducible one is again computed at NLO QCD accuracy as done for the $WZ$ case.
We then bin our events with respect to two 
angular $\phi_W$ bins, defined as $\phi \in [\pi/4,3\pi/4]$ and $\phi \in[0,\pi/4]\cup[3\pi/4,\pi]$%
, and  with respect to the $W\gamma$ system transverse mass defined as
 \bea
\left( m^T_{W\gamma}\right)^2=\left(\sqrt{m_W^2+\sum_{i=x,y}(p_i^e+\slashed p_i)^2}+p_T^\gamma\right)^2 - \sum_{i=x,y}\left(p_i^e+\slashed p_i+  p_i^{\gamma}\right )^2,
 \eea
with $m_{W\gamma}^T$ bins between [0,1000] GeV in steps of 100 GeV, [1000,1200] GeV and [1200,1500] GeV. We have chosen this variable for the binning in order to make the comparison with the $WZ$ analysis as clear as possible. By adopting this procedure we obtain the results illustrated in Fig.~\ref{fig:WA_bound} and Fig.~\ref{fig:WA_ellipses}. In Fig.~\ref{fig:WA_bound} the bounds are shown  in function of the maximum $m_{W\gamma}^T$ bin value used for the computation and for an integrated luminosity of 3000 fb$^{-1}$ assuming a systematic error of 5\%.
We can see that the dependence on the maximum $m_{W\gamma}^T$ is different for the CP-even and CP-odd operators. This is due to the fact that we can only restore the interference for the CP-even operator, due to the ambiguity in the $W$ boson decay azimuthal angle, see Eq.~\eqref{eq:nu_ambig}. We have also checked that for the obtained bounds with $m_{W\gamma}^T \lesssim 1\;$TeV the yields for the CP-even operator are dominated by the interference terms. On the other side at higher energies the quadratic terms start to dominate and the constraints on both the CP-even and CP-odd operators become similar. 
Then in Fig.~\ref{fig:WA_ellipses} we have fixed a maximum value of 1500 GeV for $m_{W\gamma}^T$ and we show the 68\% and 95\% confidence level limits. There 
 the black and red curve are computed by binning in the $\phi_W$ angle or inclusive in it respectively and where the left and right hand plot correspond assuming the SM or a BSM signal with $c_{3W}=- \tilde c_{3W}=0.3$ and we again show the bounds from the neutron and electron EDM non observation. 
As for the $WZ$ case, we can see that for the SM like signal  the binning in the $\phi_W$ angle 
practically does not change the results. This is a consequence  
of \emph{modulation from cuts} effect described in the previous Section, since the hard cut 
on the $M_T^W$ in combination with a high $p_T$ of the photon automatically select the value of the $W$ decay azimuthal angle to be close to $\pi/2$. Moreover  we can see on the right panel of 
Fig.~\ref{fig:WA_ellipses} that even in the case of assuming an injected signal, the results remain the 
same with and without the azimuthal angle binning unlike in the $WZ$ case.  As expected from Eq.~\eqref{eq:idealmodWA} and Eq.~\eqref{eq:nu_ambig} the analysis can differentiate the sign of the CP-even interaction $c_{3W}$ but it is  insensitive to the sign of the CP-odd $\tilde c_{3W}$ coupling.
In the lower panel of the Fig.~\ref{fig:WA_ellipses} we can  show the bounds obtained by including only the linear term in the production cross section Eq.\ref{eq:xsecdef}.
As expected the bounds are blind to $\tilde c_{3W}$ but for $c_{3W}$we can see that the bounds are similar to the ones obtained by the ''quadratic'' analysis.



\section{Bounds from EDMs}\label{sec:EDM}

The CP-odd operator $\tilde{O}_{3W}$ of Eq.~\eqref{eq:operators} gives also a one-loop contribution to the neutron and electron EDMs. 
Since there are strong constraints from the non observation of EDMs of elementary particles, these null measurements can potentially lead to tight bounds on $\tilde c_{3W}$. 
In particular the effective operator 
\bea
O_\gamma = i e \frac{\tilde \lambda_\gamma}{M_W^2}W^+_{\lambda\mu}W^{-,\mu}_\nu\tilde F^{\nu\lambda}
\eea
generates the EDM operator for a fermion $\psi$ 
\bea
\label{eq:OEDM}
O_{\rm EDM} =\frac{ d_f}{2}\bar \psi \sigma_{\mu\nu }\tilde F^{\mu\nu} \psi,\qquad 
\eea 
where 
\bea
d_f=\frac{g^2 e\tilde \lambda_\gamma }{64\pi^2 M_W^2}m_\psi
\eea
see {\emph{i.e.}}~\cite{Boudjema:1990dv,Gripaios:2013lea}. 
For the case of the neutron we use the form factors of~\cite{Dib:2006hk} and we obtain
\bea
d_n\simeq (1.77 d_d-0.48 d_u-0.01 d_s)\simeq 1.3 \tilde \lambda_\gamma \times 10^{-23}e\;{\rm{cm}}.
\eea
By using the latest result reported in the particle data group~\cite{Patrignani:2016xqp}, namely $|d_n|<0.3\times 10^{-25}~ e \;{\rm{cm}}$  at 90\% CL, we obtain a limit
\bea
|\tilde \lambda_\gamma|\lesssim 0.0023
\eea
which translates in
\bea
\left|\frac{\tilde c_{3W}}{{\rm{TeV}}^2}\right|\lesssim \frac{0.36}{{\rm TeV}^2}
\eea
 which is of the same order  with the bounds attainable at the end of the HL-LHC phase from the precision measurements of the $W\gamma$ and $WZ$ processes.
On the other side the experimental limit on the electron EDM is much stronger than the one of the neutron, $|d_e|<0.87\times 10^{-28}e\;{\rm cm}$ at  90\% CL~\cite{Patrignani:2016xqp}. This leads to a much stronger constraint on the Wilson coefficient of the CP violating triple gauge coupling operator. Namely we obtain
\bea
|\tilde \lambda_\gamma|\lesssim 8.3\times 10^{-5},
\eea
which implies
\bea
\left|\frac{\tilde c_{3W}}{{\rm{TeV}}^2}\right|\lesssim \frac{0.013}{{\rm TeV}^2}
\eea which is far beyond the reach of current and future collider experiments.

We stress however that these bounds can potentially be relaxed in presence of additional new physics contribution affecting the $O_{\rm EDM}$ operator of Eq.~\eqref{eq:OEDM} and cancelling 
 against the one-loop contribution arising from $O_{3\tilde W}$. We don't discuss this possibility any further, stressing again that the limits arising from the non observation of an electron EDM are potentially more constraining that the ones arising from direct LHC measurements.

\section{High Energy LHC}\label{sec:LHC27}

By the end of 2035 the LHC experiments ATLAS and CMS will have collected $\sim3\;$ab$^{-1}$ of integrated luminosity each, ending the HL phase of the CERN machine. Various collider prototypes have been proposed in the recent years for the post LHC era. These include leptonic machines such as ILC and CLIC ideal for performing precision measurements of the Higgs couplings, and hadronic machines, as the FCC-hh, a 100 TeV proton-proton collider, with huge potentiality for the discovery of resonant new physics above the TeV scale,  that however requires enormous efforts, among which a new $\sim 100\;$Km tunnel.
Hence in the last years a lot of attention has been given to the possibility of building a higher energy proton collider within the LHC tunnel. Thanks to new techniques with which it would be possible to build 16 T magnets, a centre of mass energy of 27 TeV can be envisaged. This doubling of energy with respect to the LHC can offer great physics opportunities~\cite{CidVidal:2018eel} both for on-shell particle productions, but also for indirect measurements as the ones discussed in this paper. 

\begin{figure}[t!]
\centering
{\includegraphics[width=0.46\textwidth]{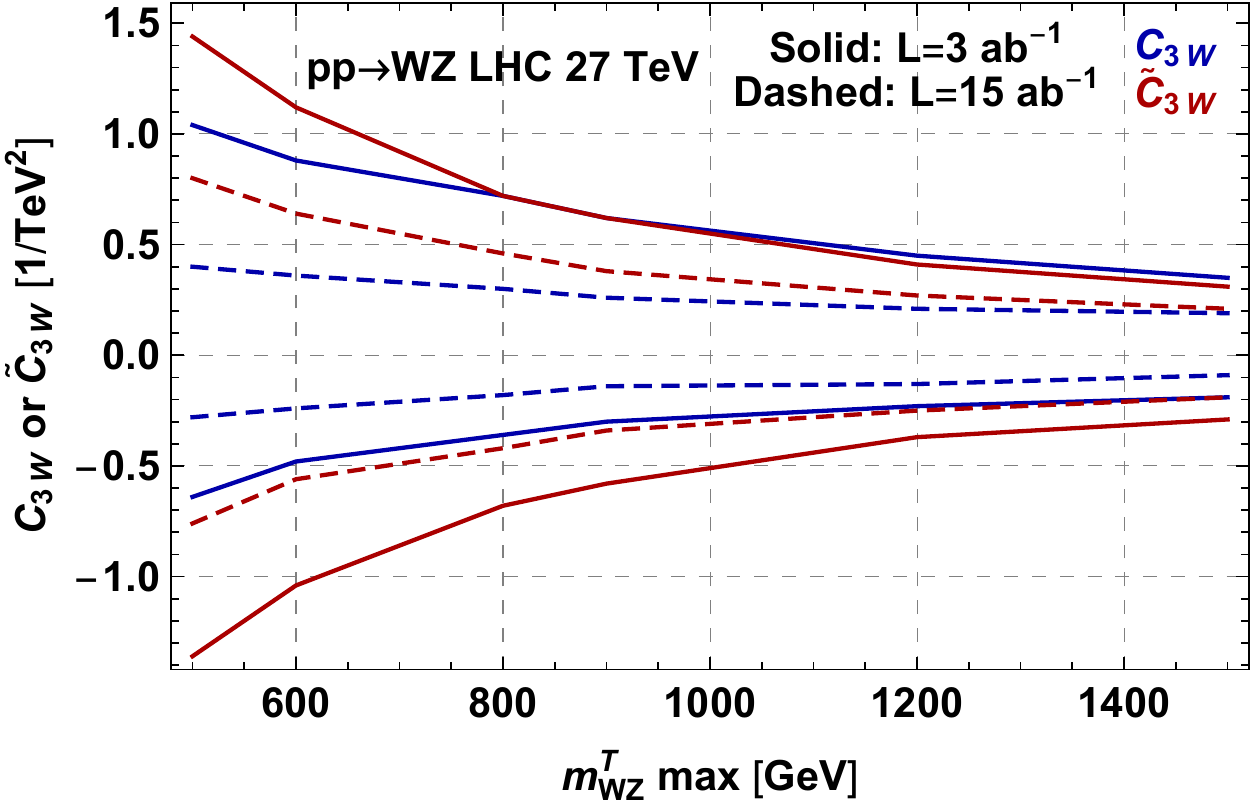}}\hfill
\caption{95\%  bound on the $c_{3W}$ and $\tilde c_{3 W}$ Wilson coefficients computed with four and two equally spaced angular bins for $\phi_Z$ and $\phi_W$ respectively, in function of the largest $WZ$ system transverse mass bin used for the 27 TeV LHC with 3 ab$^{-1}$ (solid) and 15 ab$^{-1}$ (dashed) of integrated luminosity.  A systematic error of 5\% has been assumed.  }
\label{fig:WZ_bound_27}
\end{figure}

\begin{figure}[t!]
\centering
\includegraphics[scale=0.55]{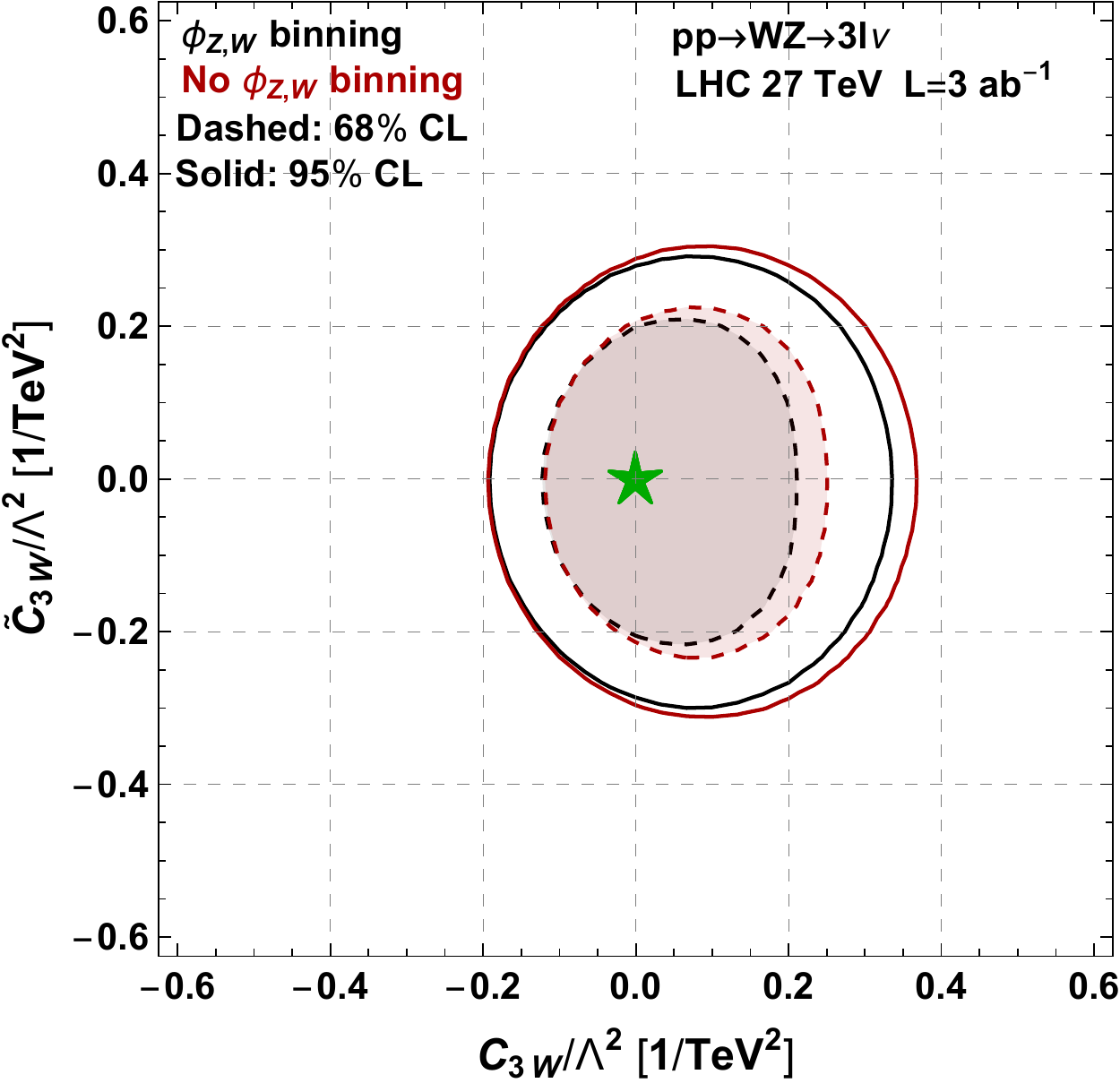}\hfill
\includegraphics[scale=0.55]{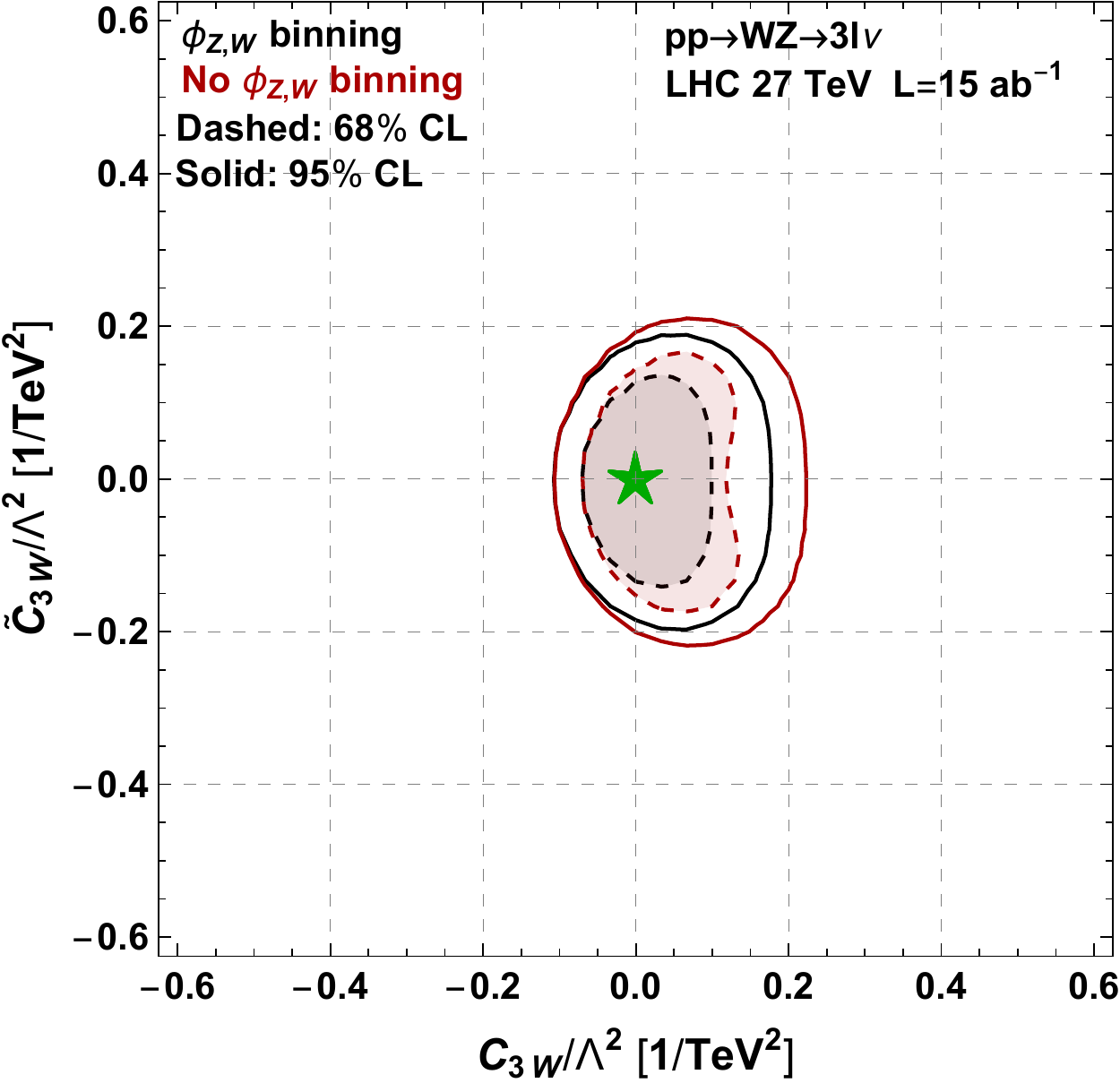}\\
\vskip 5mm
\includegraphics[scale=0.55]{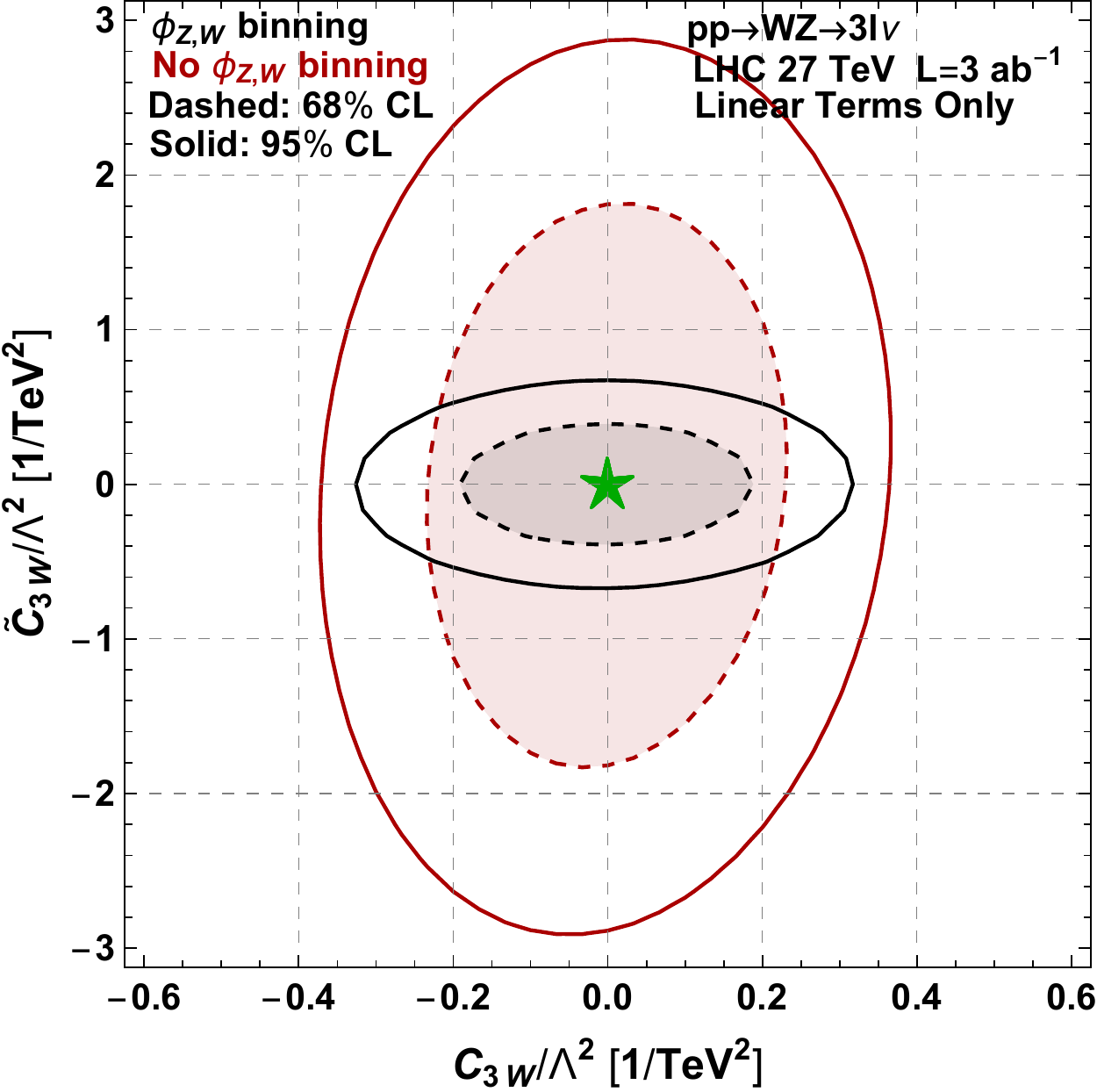}\hfill
\includegraphics[scale=0.55]{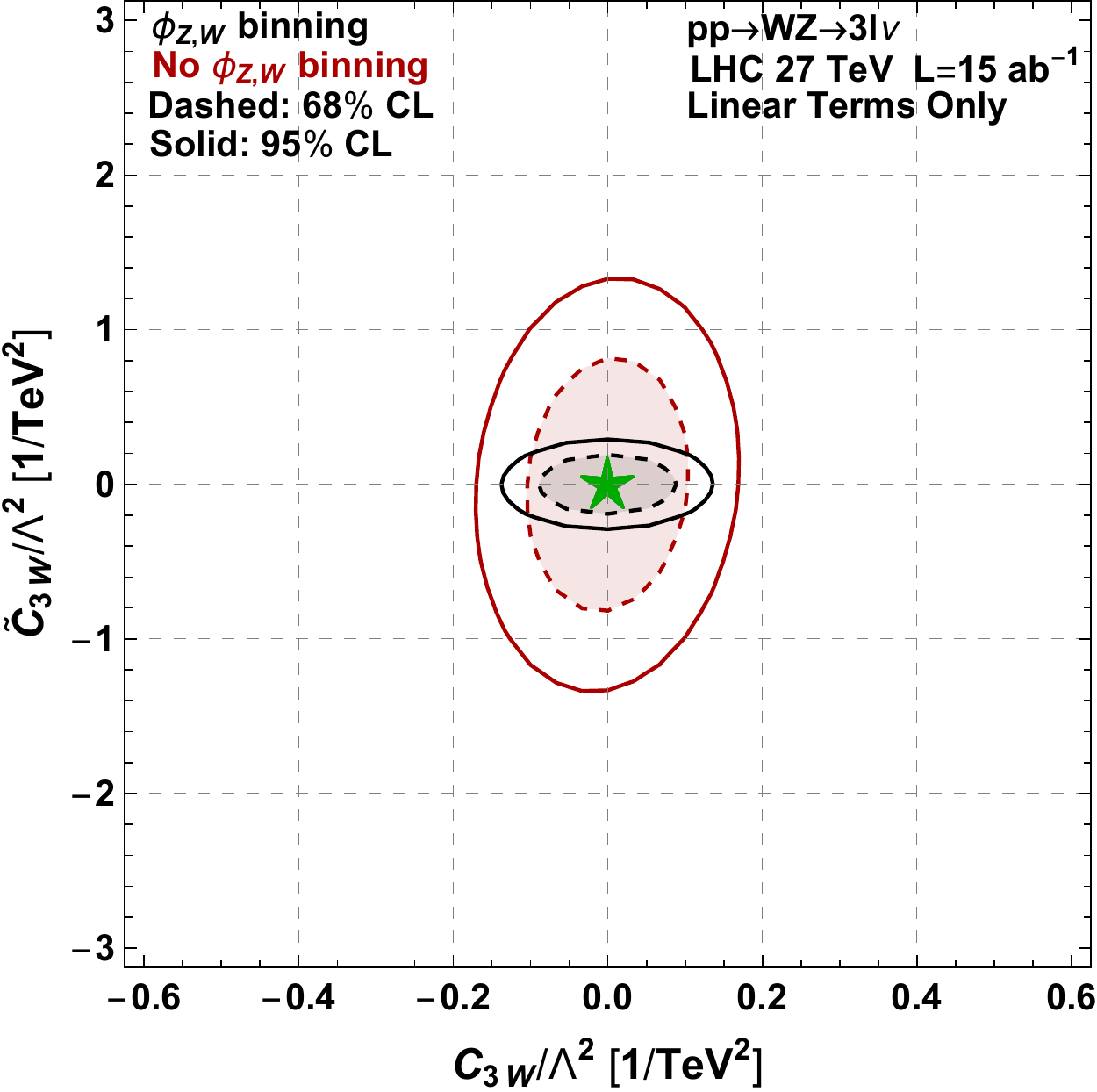}
\caption{ 68\%  (dashed) and 95\%  (solid) posterior probability contours for the $WZ$ at 27 TeV analysis with (black) and without (red) the binning in the $\phi_Z$ and $\phi_W$ angles, see main text for more details. Only events with $m^T_{WZ} <1.5 $ TeV are used. 
The upper and lower panels correspond to the limits obtained with and without the inclusion of the quadratic term in the EFT expansion respectively.
}
\label{fig:WZ_ellipses_27}
\end{figure}

\begin{figure}[t!]
\centering
{\includegraphics[width=0.46\textwidth]{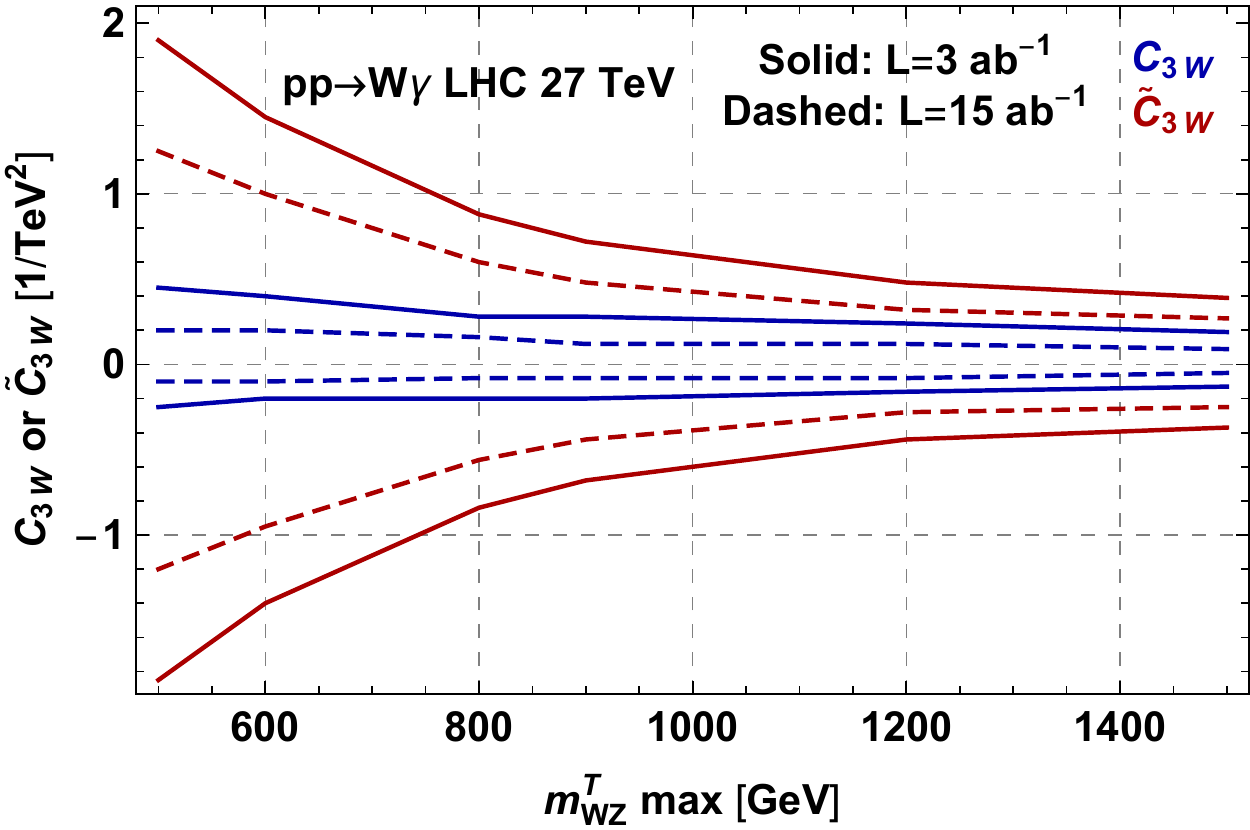}}\hfill
\caption{95\%  bound on the $c_{3W}$ and $\tilde c_{3 W}$ Wilson coefficients computed with two equally spaced angular $\phi_W$ bins in function of the largest $W\gamma$ system transverse mass bin used for the 27 TeV LHC with 3 ab$^{-1}$ (solid) and 15 ab$^{-1}$ (dashed) of integrated luminosity.  A systematic error of 5\% has been assumed.  }
\label{fig:WA_bound_27}
\end{figure}

\begin{figure}[t!]
\centering
\includegraphics[scale=0.55]{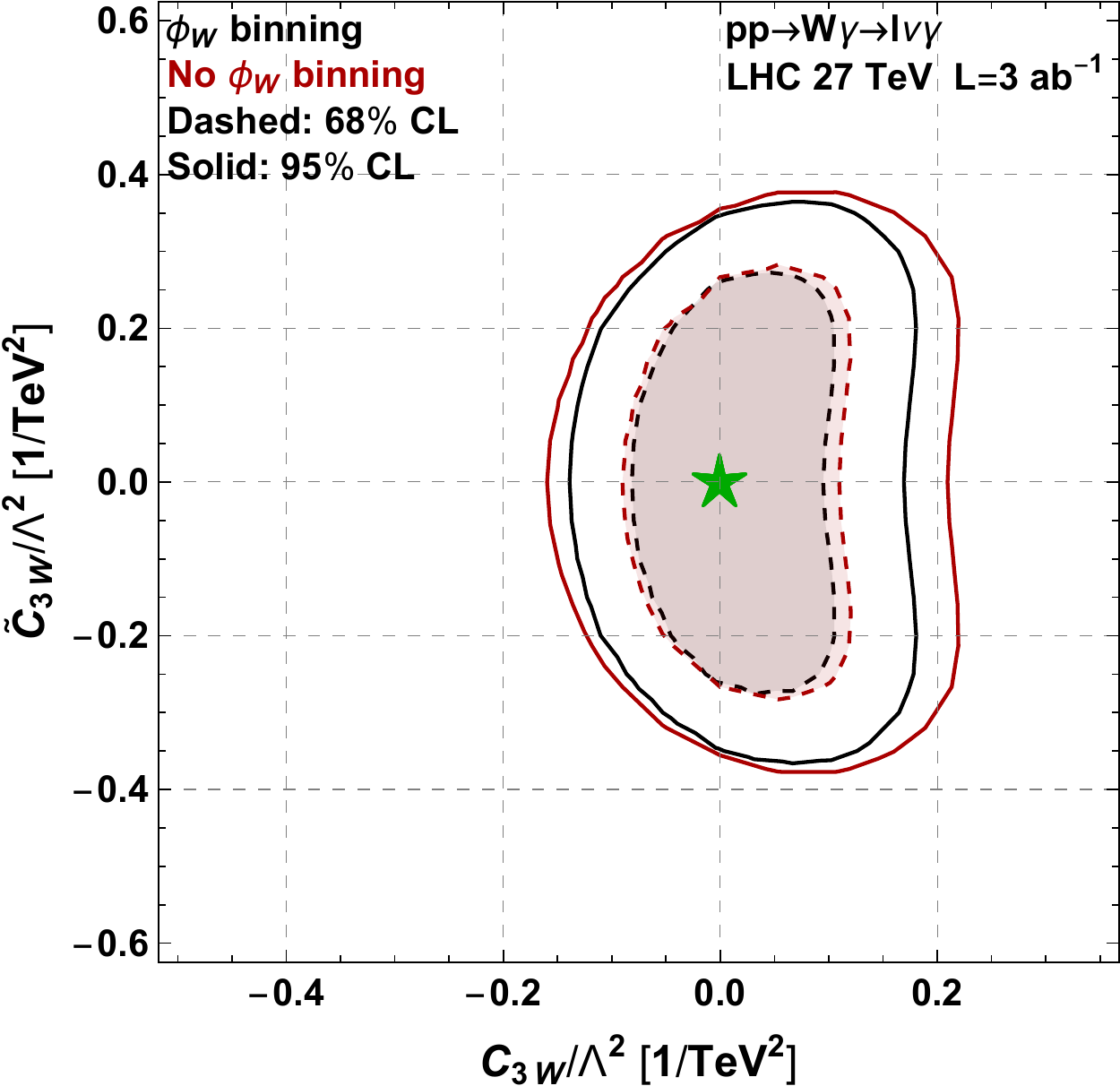}\hfill
\includegraphics[scale=0.55]{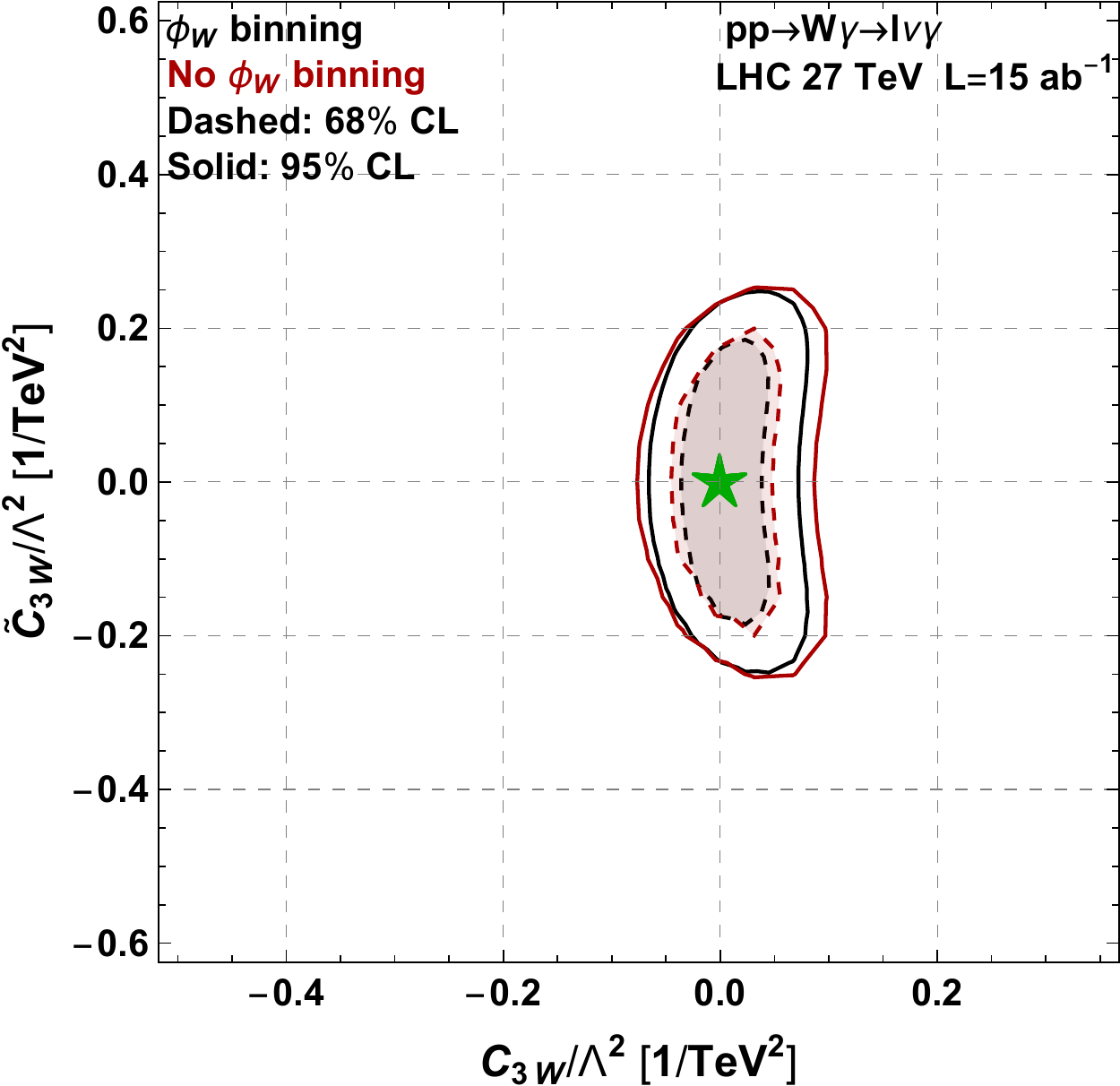}\\
\vskip5mm
\includegraphics[scale=0.55]{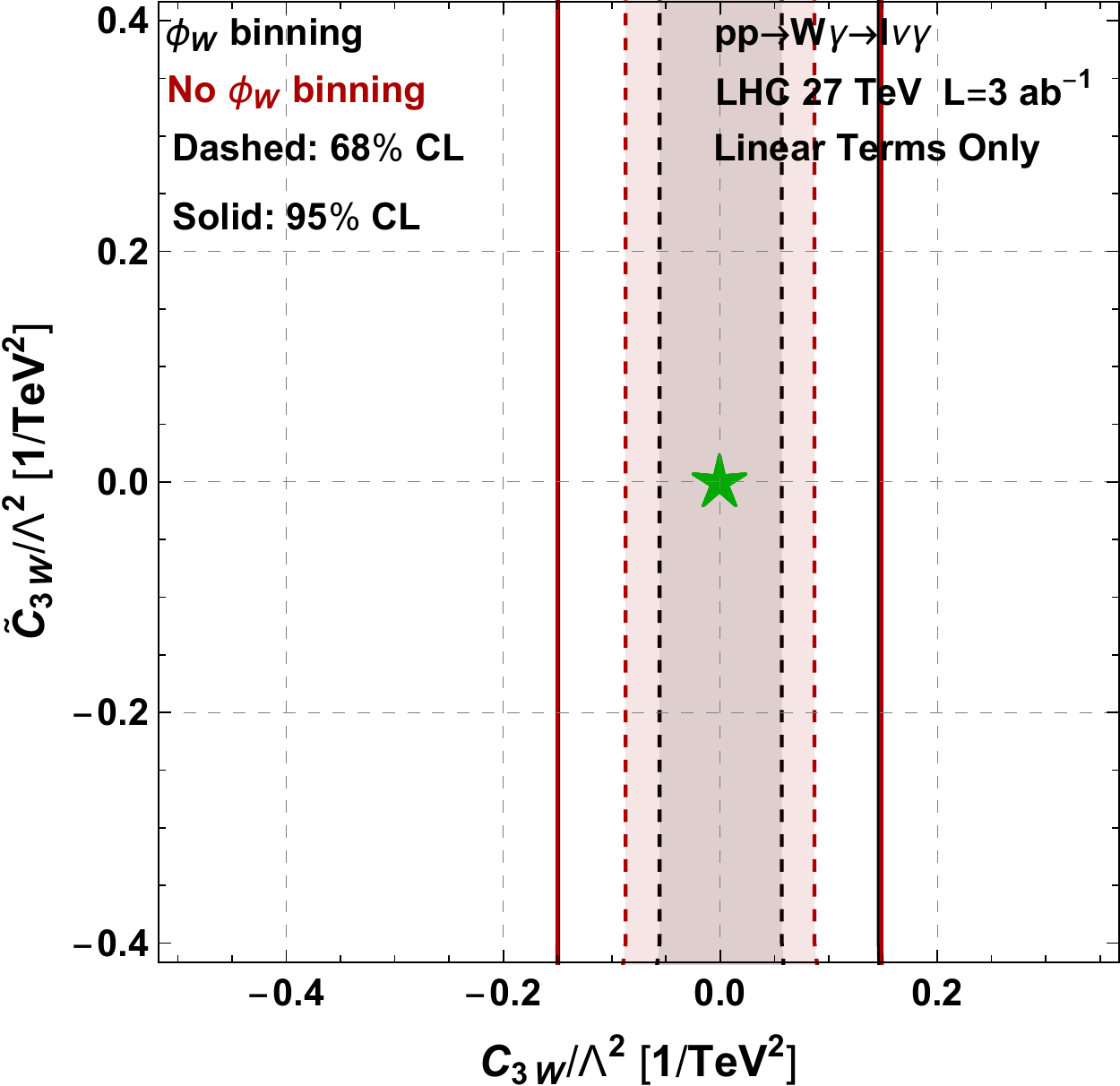}\hfill
\includegraphics[scale=0.55]{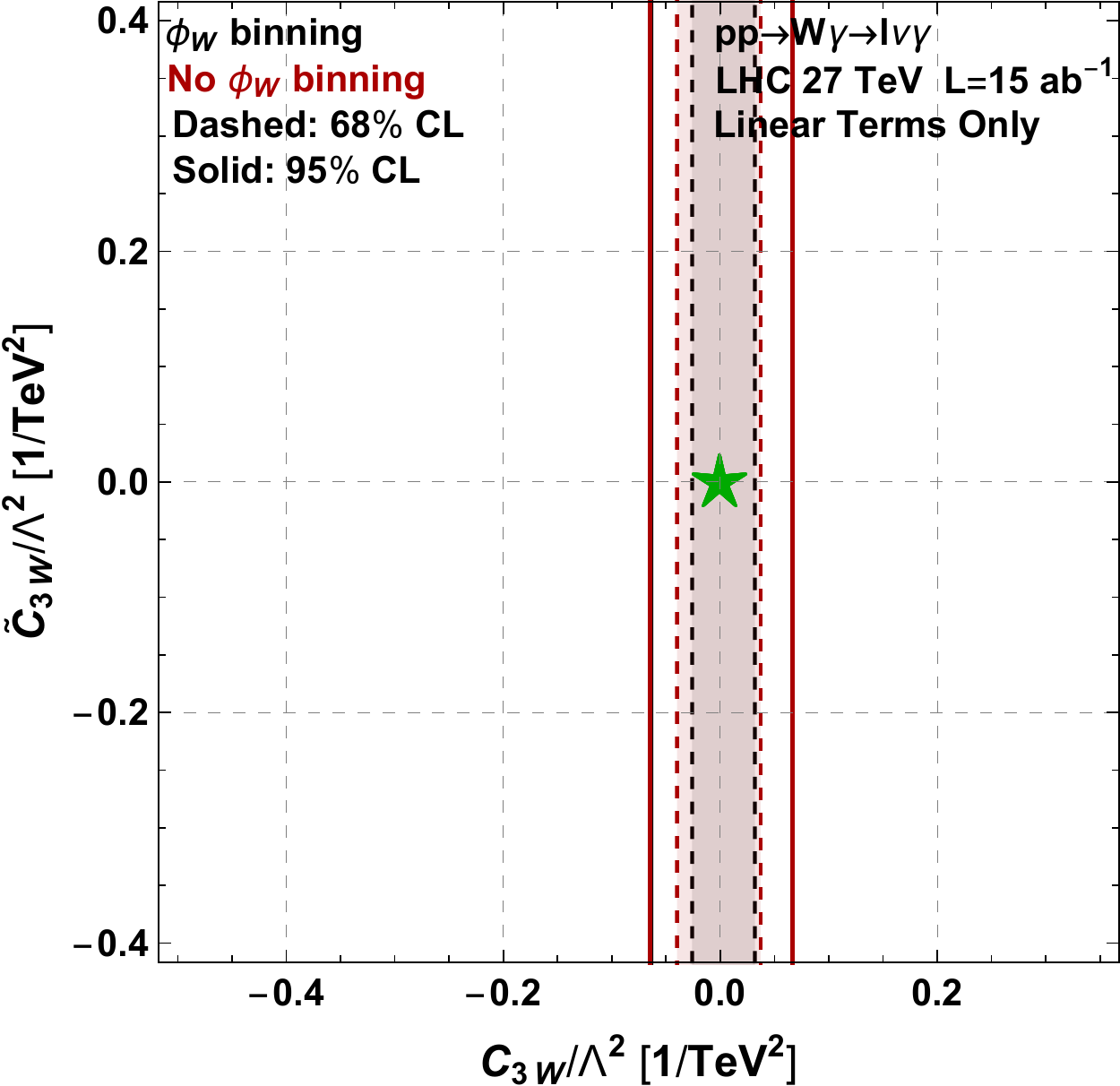}\\
\caption{ 68\%  (dashed) and 95\%  (solid) posterior probability contours for the $W\gamma$ at 27 TeV analysis with (black) and without (red) the binning in the $\phi_W$ angle, see main text for more details. Only events with $M_T^{W\gamma} <1.5 $ TeV are used. The upper and lower panels correspond to the limits obtained with and without the inclusion of the quadratic term in the EFT expansion respectively.
}
\label{fig:WA_ellipses_27}
\end{figure}

We then show in this Section the prospects for measuring the $c_{3W}$ and $\tilde c_{3W}$ Wilson coefficients by applying   analyses similar to the ones discussed in Sec.~\ref{sec:WZ} and Sec.~\ref{sec:WA}. We focus on two benchmark of integrated luminosites: 3 ab$^{-1}$ and 15 ab$^{-1}$.
The results are shown in Fig.~\ref{fig:WZ_bound_27}-Fig.~\ref{fig:WA_ellipses_27}, in complete analogies with the figures of the previous Sections.

For the $WZ$ analysis we can see that the relative  improvement   from the binning in $\phi_Z$ and $\phi_W$ angles increases compared to the $14$ TeV analysis, since we are getting closer to the values of the Wilson coefficients when the interference term dominates the cross section. Similar effects hold for the $W\gamma$ process. The effect of the modulation from cuts becomes less important since for the same values of the $m^T_{W\gamma}$ variable, larger values of the longitudinal momentum are expected at higher collision energies, so  that the $\phi_W\sim \pi/2$ bin selection becomes less strong,  see discussion in the Section \ref{sec:mod_cuts}.

\section{Summary}\label{sec:concl}
We have analyzed the diboson production,  $pp\to WZ$ and $pp\to W\gamma$ at NLO QCD order in the 
presence of the dimension six operators of Eq.~\eqref{eq:operators}, paying a particular 
attention to the effects related to the interference between the SM and BSM contributions.
We have found that NLO QCD effects mildly affects the results of the analogous LO analysis~\cite{Azatov:2017kzw}, since the helicity selections rules do not apply at NLO.
For both the  $pp\to WZ$ and $pp\to WZ,W\gamma$ processes  the observables related to the azimuthal angles lead to an enhancement of the interference providing a better sensitivity to the 
new physics interactions. In order to estimate the LHC possibilities on measuring these interaction we have closely followed available experimental studies of  diboson production ~\cite{Chatrchyan:2013fya,ATLAS:2018ogj}.
 Interestingly we  have found that some of the kinematic selection cuts needed  to suppress the reducible backgrounds in realistic analyses  are partially performing an azimuthal angular bin selection.  This effect turns out to be  particularly important for the  $pp\to W\gamma$ processes where the strong cut on the $M_W^T$ forces the azimuthal angle to be close to $\pi/2$, making a further binning in the azimuthal angle $\phi_W$ less important with respect to what is naively expected.
The prospects of the bounds at the HL  and HE phases of the LHC are 
presented. This leads to a sensitivity $\sim 10^{-3}$ on 
the triple gauge couplings $\lambda_Z$ and $\tilde \lambda_Z$ at HL-LHC~\footnote{The coefficient $\lambda_Z$ is normalized as $O_{\lambda_Z} = \lambda_Z \frac{i g}{m_W^2} W^{+\mu_2}_{\mu_1} W^{-\mu_3}_{\mu_2}W^{3\mu_1}_{\mu_3}$. An analogous definition holds for $\tilde \lambda_Z$.}.  The   HE phase of the CERN machine can further improve the bounds by factor of $\sim2-5$. These results are summarized in Tab.~\ref{tab:results_final}.

\begin{table}[!h]
\centering
\begin{tabular}{c|c|c|c c | c c}
\multirow{2}{*}{Channel} &  \multirow{2}{*}{Energy} &  \multirow{2}{*}{Luminosity} & \multicolumn{2}{c|}{$\lambda_Z~[\times 10^{-3}]$} & \multicolumn{2}{c}{$\tilde \lambda_Z~[\times 10^{-3}]$} \\
 & & & 68\%  & 95\%  & 68\%  & 95\% \\
\hline\hline
\multirow{ 3}{*}{$\bm{WZ}$} & 14  TeV& 3 ab$^{-1}$  & [-2.1,\;1.2] & [-2.9,\;1.7] & [-1.7,\;1.7] & [-2.4,\;2.4]\\
& \multirow{ 2}{*}{27 TeV} & 3 ab$^{-1}$  & [-1.4,\;0.7] & [-2.2,\;1.2] & [-1.5,\;1.3] & [-2.0,\;1.8]\\ 
 &  & 15 ab$^{-1}$ & [-0.7,\;0.4] & [-1.2,\;0.6] & [-0.9,\;0.8] & [-1.3,\;1.2]\\ 
 \hline
\multirow{ 3}{*}{$\bm{W\gamma}$} & 14  TeV& 3 ab$^{-1}$ & [-1.2,\;0.9] & [-2.0,\;1.6] & [-2.2,\;2.1] & [-3.0,\;2.9]\\
& \multirow{ 2}{*}{27 TeV} & 3 ab$^{-1}$ & [-0.7,\;0.4] & [-1.2\;0.8] & [-1.8,\;1.7] & [-2.5,\;2.4] \\ 
 &  & 15 ab$^{-1}$ & [-0.4,\;0.2] & [-0.6,\;0.3] & [-1.3,\;1.2] & [-1.7,\;1.5] \\ 
\end{tabular}
\caption{Summary of the results for the various channels  in terms of the CP-even and CP-odd anomalous triple gauge couplings. Only events with $m^T_{WZ, W\gamma} <1.5 $ TeV are used.}
\label{tab:results_final}
\end{table}

\subsection*{Acknowledgement}
We would like to thank J. Elias-Miro for the collaboration on the initial stages of the project and K Mimasu and B. Fuks for providing us help with the model {\tt HELatNLO\_UFO} .

\bibliographystyle{JHEP}

{\footnotesize
\bibliography{biblio}}

\end{document}